\documentclass[journal=jctcce,manuscript=article]{achemso}
\usepackage{amssymb,amsmath,amsfonts,multicol,multirow,longtable,array,mathpazo}
\usepackage{lscape}
\usepackage[version=3]{mhchem} % Formula subscripts using \ce{}
\usepackage{appendix}
\usepackage{soul} %text highlighting
\usepackage[usenames]{color}
\usepackage{makecell}
\usepackage{enumerate}
\usepackage{xr}
\usepackage[T1]{fontenc} % Use modern font encodings
\usepackage{booktabs}
\usepackage{tikz}
\usetikzlibrary{shapes.geometric}
\usepackage{threeparttable}
\usepackage{graphicx}
\usepackage{subfigure}
\usepackage{colortbl}
\usepackage{framed}
\usepackage{verbatim}
\usepackage{amsbsy}
\usepackage{bm}% bold math
\usepackage{bbm}
\usepackage[normalem]{ulem}
\usepackage{float}
\usepackage{algorithm}
\usepackage{algpseudocode}
\usepackage{float}
\usepackage{subfigure}
\usepackage{listings}
\newcounter{xscheme}

\newfloat{Algorithm}{htbp}{alg}
\floatname{Algorithm}{Algorithm}

\newcounter{exe}[figure]
\newcommand{\iexe}{\refstepcounter{exe}\the\value{exe}:}

\setkeys{acs}{maxauthors = 0} % references of many authors

\author{Ning Zhang}
\author{Qingpeng Wang}
\author{Wenjian Liu}\email{liuwj@sdu.edu.cn}
\affiliation{Qingdao Institute for Theoretical and Computational Sciences and Center for Optics Research and Engineering,
	Shandong University, Qingdao 266237, China}

\title{\texttt{MetaWave}: A Platform for Unified Implementation of Nonrelativistic and Relativistic Wavefunctions}

\setkeys{acs}{articletitle=true}

\begin{document}
	
\begin{abstract}
\texttt{MetaWave} is a C++ template-based architecture designed for unified implementation of  
nonrelativistic and relativistic wavefunction-based quantum chemical methods.
It is highly modular, extendable, and efficient.
This is achieved by decoupling the three distinct aspects of quantum chemical methods
 (i.e., nature of Hamiltonian, structure of wavefunction, and strategy of parallelization ),
thereby allowing for separate treatment of them through their internal type-trait and tagging systems furnished by C++ metaprogramming.
Once the second-quantized Hamiltonians, whether nonrelativistic (spin-free) or relativistic (spin-dependent),
are decomposed into topologically equivalent diagrams for a unified evaluation of the basic coupling coefficients between (randomly selected)
spin-free or spin-dependent configuration state functions or Slater determinants incorporating
full molecular symmetry (including single or double point group and spin or time reversal symmetry),
the many-electron wavefunctions, whether built up with scalar or spinor orbitals, can be assembled with the same templates.
As for parallelization, \texttt{MetaWave}
supports both OpenMP and MPI,
with the majority of the latter being translated automatically from its OpenMP counterparts.
The whole structure of \texttt{MetaWave} is reviewed here, with some showcases for illustrating its performance.
\end{abstract}

\maketitle

\clearpage
\newpage

\section{Introduction}
As emphasized recently\cite{LiuWIRES2023}, high-precision spectroscopic calculations ought to
treat relativistic, correlation, and quantum electrodynamics (QED) effects on an equal footing, so as to match experimental measurements as closely as possible.
Since both relativistic and QED effects can be incorporated into second-quantized Hamiltonians\cite{eQED,LiuPhysRep,LiuPerspective2020,CommentQED}, which
take the same form as the nonrelativistic (spin-free) one, it would be nice if all second-quantized Hamiltonians can be manipulated
in the same way for subsequent development of wavefunction methods for describing electron correlation.
This is indeed possible, thanks to two recent developments: (1) relativistic-QED Hamiltonians, whether four-component or two-component,
can be constructed in a unified manner based solely on physical arguments\cite{UnifiedH}; (2)
all second-quantized Hamiltonians can be decomposed into topologically
% the same
equivalent
diagrams, such that
 the basic coupling coefficients between (randomly selected)
spin-free or spin-dependent configuration state functions (CSF) or Slater determinants (DET) can be evaluated in a unified
manner\cite{4C-iCIPT2}.
What is left is then how to design and implement many-electron wavefunctions for describing electron correlation as accurately as possible.
The so-called strongly correlated systems of electrons, such as extended $\pi$ systems and polynuclear transition metal clusters,
are particularly intricate in this aspect, due to the fact that the static and dynamic components of the overall correlation
are therein strongly entangled and even interchangeable. The currently best methods for such kinds of systems
include density matrix renormalization group (DMRG)\cite{Schollwock2011density,DMRGb,ChanLowEntanglement2012,DMRG2015,Wouters2014,DMRG_Legeza_2,DMRG_Reiher_Review} and
selected configuration interaction (sCI) plus second-order perturbation theory (sCIPT2)\cite{iCI,iCIPT2,iCIPT2New,iCISCF,HBCI2016,HBCI2017a,
HBCI2017b,HBCI2017c,SHBCI2018,HBCICASSCF,
HBCIOrbOpt2021,Proj-sCI2016,ACI2016,ACI2017,ACI2018,ASCI2016,ASCI2017,ASCI2018PT2,ASCISCF,ICE-1,ICE-2,
4C-iCIPT2,SOiCI,SOiCISCF,HBCISOC,4CSHCI}.

Given so many wavefunction methods, the question is how to extend them to the relativistic realm, yet without significant code rewriting. 
To address this, we first outline the general aspects of wavefunctions methods:

\iffalse
Even for a relatively simple post-Hartree–Fock method like second-order Møller–Plesset perturbation theory (MP2), implementation is far from straightforward.
Consider the canonical MP2 correlation energy expression:
\begin{equation}
	E_{\mathrm{c}}^{\mathrm{MP} 2}=\sum_{i, j, a, b}-\frac{1}{4} \frac{|\langle i j \| a b\rangle|^2}{\Delta_{i j}^{a b}}, \label{MP2}
\end{equation}
where
$i,j$ are used for occupied orbitals and $a,b$ are used for unoccupied orbitals respectively.
The term $\langle i j \| a b\rangle = \langle i j | a b\rangle-\langle i j | b a\rangle$ denotes the antisymmetrized two-electron repulsion integrals and the denominator $\Delta_{i j}^{a b}=\varepsilon_a+\varepsilon_b-\varepsilon_i-\varepsilon_j$ is computed from the corresponding orbital energies $\epsilon$.
Conceptually, implementing canonical MP2 involves simple looping over four indices, but practical challenges quickly arise:
\fi
\begin{enumerate}
	\item \textbf{Nature of Hamiltonian}\\
To implement a wavefunction method, the very first thing under concern is the nature of the chosen Hamiltonian,
	spin-free or spin-dependent. While
molecular orbitals can be chosen to be real-valued in the spin-free case, they
are necessarily complex-valued in the spin-dependent case.
This leads to different permutation and point group symmetries to molecular integrals
and hence different data structure for their storage.

	\item \textbf{Structure of Wavefunction}\\
$N$-electron wavefunctions can be expanded in terms of Hilbert space basis states (HSS) or Fock space basis states (FSS).
% Fock space means the full Hilbert space across all particle number sectors.
Every member of the former carries $N$-electrons and $N$ spin orbitals, whereas that of the latter may have a variable number of electrons
and even a variable number of spin orbitals.
For instance, the well-established many-body perturbation theory (MBPT), coupled-cluster (CC), and configuration interaction (CI) types of wavefunctions
are all built up with an $N$-electron mean-field (Hartree-Fock or CASSCF) reference and additional
$N$-electron configurations generated from it by systematic excitations/replacements.
No matter how such HSSs are herein truncated, antisymmetry is always maintained automatically.
In contrast, when working with \emph{local} FSSs as in DMRG wavefunctions, suitable constraints must be
imposed from the outside to conserve the number ($N$) of electrons and antisymmetry.
In essence, the two kinds of representations access electron correlation through different routes:
the HSS representation tries to improve the initial independent-particle description by selecting
low-energy $N$-electron functions from the (first-order) interacting Hilbert space, ending up with a linear (in CI) or nonlinear (in CC) combination of limited HSS,
whereas the local FSS representation attempts to improve the initial independent-subsystem description by
%renormalizing the local states iteratively.
selecting high-entanglement local states from the renormalized Fock space,
ending up with a particular linear combination of \emph{all} HSS.
%encoding entanglements between the subsystems via a renormalization procedure (i.e., contraction of tensors).
% correlations are introduced between the orbital Fock spaces, by contracting the site-functions (tensors)
The symmetry adaptation of HSS
can be achieved by using appropriate irreducible representations (irrep) of the $U(n)$ and single point groups
in the spin-free case, whereas by using time reversal and double point groups in the spin-dependent case.
In contrast, the symmetry adaptation of local FSS has to be performed in terms of local symmetry operations
in both the spin-free\cite{SADMRG,SA-MPS-MPO} and spin-dependent\cite{4C-DMRG-JCP2018} cases.
Additionally, the implementation has to distinguish between complex and real algebras
associated with spin-dependent and spin-free wavefunctions, respectively.
	
\item \textbf{Strategy of Parallelization}\\
The efficient execution of wavefunction methods relies on proper parallelization and hardware-specific optimizations. Different parallelization patterns
(e.g., OpenMP or MPI) or hardware platforms (e.g., CPU,  GPU or their hybrid\cite{CPU-GPU-DMRG-2024,CPU-GPU-CC-2024}) typically necessitates substantial code rewrites.
%Consequently, the entire program must be adapted not only to the parallel strategy but also to the concurrent new memory management and data communication patterns, thereby complicating the portability and scalability.
\end{enumerate}
To achieve a unified implementation of wavefunction methods, the above aspects must be decoupled as much as possible. To this end,
we design here a platform, \texttt{MetaWave}, which can treat the three aspects separately through their internal type-trait\footnote{Type traits are compile-time tools in C++ that provide information about types (e.g., inquiry whether a type represents the nonrelativistic Hamiltonian). This enables functions or templates to automatically adjust their behavior based on the types they handle, enhancing flexibility and type safety.} and tagging \footnote{
	A tag in C++ is an empty struct to label concepts.
	For example, one can use a struct \lstinline|NonRelativistic| to label nonrelativistic Hamiltonian.
	Combined with type-traits, this will enable compilers to process information related to the nonrelativistic Hamiltonian.
}
systems furnished by C++ metaprogramming. What is essential is the use of a template to generate temporary subroutines,
which are then merged with other relevant codes and compiled.
This strategy reduces significantly code duplication and enhances maintainability as the software expands.
Use of \texttt{MetaWave} has been made to
achieve a unified implementation\cite{4C-iCIPT2} of the nonrelativistic\cite{iCIPT2,iCIPT2New} and relativistic\cite{SOiCI,4C-iCIPT2} variants of the
fully symmetry adapted iCIPT2 (iterative configuration interaction with
selection and second-order perturbation) method. Note in passing that the idea of decoupling different aspects of 
wavefunction methods was already adopted in the HORTON program package\cite{HORTON}, where the Fanpy library\cite{FanpyJCC2023}
is also highly modular and general for implementing new wavefunction methods based on the
flexible ans\"atze for $N$-electron configuration interaction (FANCI)\cite{FANCI} and perturbation theory (FANPT) frameworks\cite{FANPT}. 
However, Fanpy is a research tool written purely in Python,
with the primary goal being quick implementation and test of a new wavefunction ansatz, instead of high-performance parallel computing.
Moreover, Fanpy does not support relativistic Hamiltonians and wavefunctions, which are the primary concern of \texttt{MetaWave}.

The remaining of the paper is organized as follows.
Sec. \ref{Background}
is devoted to general features of the sCIPT2 methods,
including unified treatments of Hamiltonians, many-particle basis functions (MPBF), and Hamiltonian matrix elements (HME), selection procedures,
and proper implementation of the Epstein-Nesbet second-order perturbation theory (ENPT2)\cite{Epstein,Nesbet}.
The three-layer architecture of \texttt{MetaWave} is then presented in detail in Sec. \ref{architecture}.
As a showcase, the computational costs of sf-X2C-iCIPT2, SOiCI, and 4C-iCIPT2 are briefly compared in Sec. \ref{Application}.
The paper is finally closed with concluding remarks in Sec. \ref{Conclusion}.

\section{sCIPT2} \label{Background}

\begin{figure}
	\includegraphics[width=0.7\textwidth]{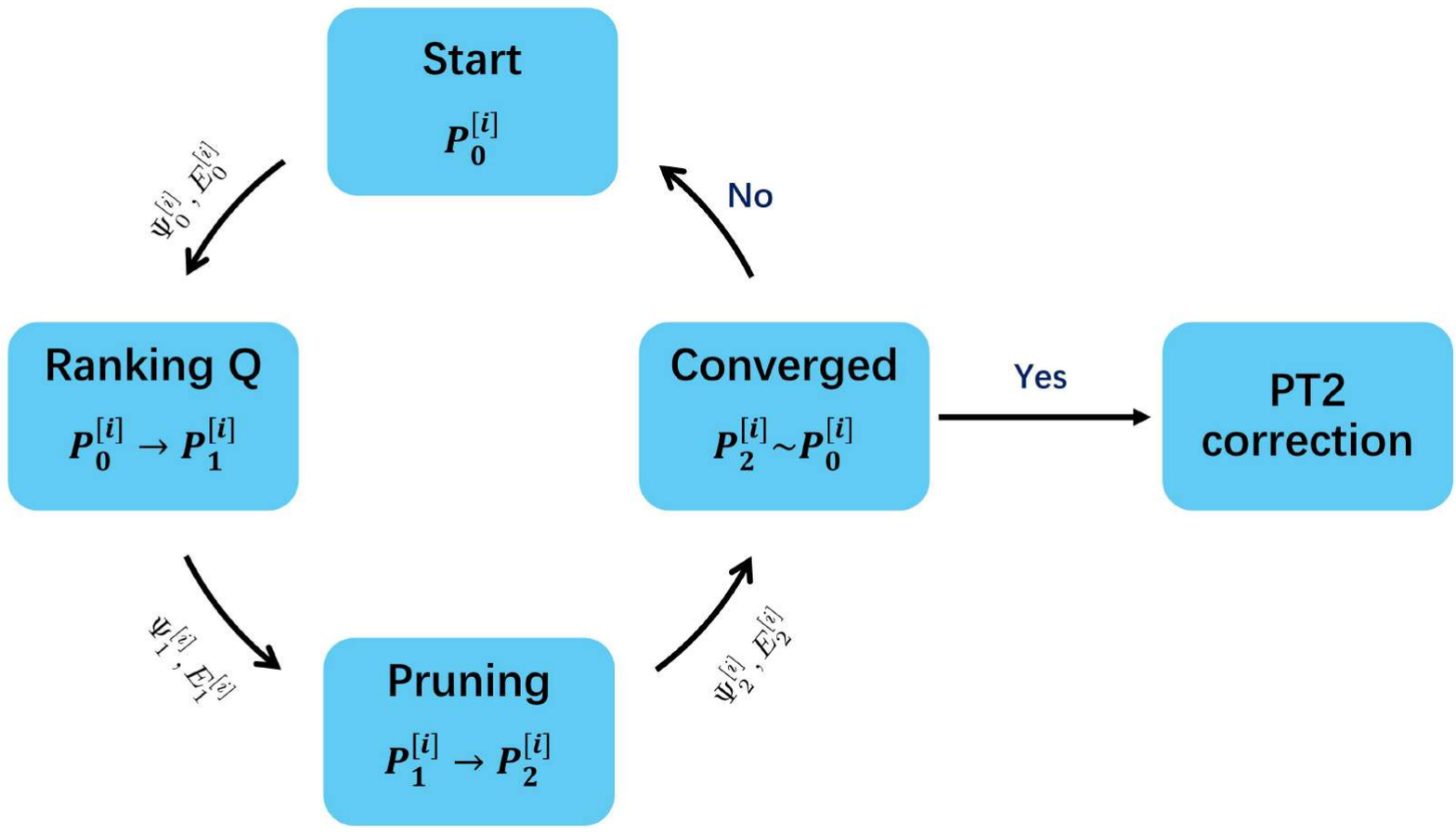}
	\caption{Workflow of sCIPT2}
	\label{sCI_workflow}
\end{figure}

The available nonrelativistic\cite{iCI,iCIPT2,iCIPT2New,iCISCF,HBCI2016,HBCI2017a,HBCI2017b,HBCI2017c,SHBCI2018,HBCICASSCF,
HBCIOrbOpt2021,ACI2016,ACI2017,ACI2018,ASCI2016,ASCI2017,ASCI2018PT2,ASCISCF,ICE-1,ICE-2}
and relativistic\cite{4C-iCIPT2,SOiCI,SOiCISCF,HBCISOC,4CSHCI} sCIPT2 methods implemented
in several program packages\cite{FORTE,Q-Chem,ORCA,QuantumPackage,sanshar_dice,BDF2020}
all involve three major steps, i.e., preparation of a guess space $P_0$, iterative update of the variational space until convergence,
and second-order perturbation correction to the energy.
However, they differ in the actual implementations, which involve the following issues (see Fig. \ref{sCI_workflow}):
\begin{enumerate}
	\item How to assess and rank the importance of the CSFs/DETs outside of the variational space $P_0$,
so as to expand $P_0$ to $P_1$?
	\item How to diagonalize $P_1$ in the pruning step, so as to reduce $P_1$ to $P_2$?
	\item How to terminate the selection process?
	\item How to estimate the contribution of the discarded CSFs/DETs?
\end{enumerate}
%These points also encapsulate four key components of sCIPT2 methods: the ranking criterion, the pruning process, the convergence criterion and the perturbation correction.
%Furthermore, the method to diagonalize the space $P_i$ should be also considered as a key component of sCIPT2 methods.

Herewith, we introduce a general algorithm that is suitable for different Hamiltonians, MPBFs,
and parallelization schemes.

\subsection{Unified Handling of Hamiltonians}
All Hamiltonians in ab initio quantum chemistry, whether spin-free or spin-dependent, take the following generic, second-quantized form
\begin{align}
	H&=\sum_{i, j} h_{i j} E_{i j} +\frac{1}{2} \sum_{i, j, k, l}(i j| k l) e_{i j, k l} =H^\dag \label{Hamiltonian}\\
    E_{i j}   &= a_i^\dagger a_j = E_{ji}^\dagger\\
    e_{ij,kl} &= a_i^\dagger a_k^\dagger a_l a_j =e_{kl,ij}=e_{ji,lk}^{\dagger}=e^{\dagger}_{lk,ji}\label{sym_op}
\end{align}
Here, $\{i,j,k,l\}$ refer to molecular orbitals/spinors obtained by a mean-field calculation, whereas
$h_{i j}$ and $(ij|kl)$ (in the Mulliken notation) are the integrals of the core Hamiltonian and two-body operators, respectively.
It should be emphasized that, in the spin-dependent case, the molecular spinors are assumed here to be symmetrized\cite{Symm2009}
according to both double point group and time reversal symmetries, such that they can be grouped into $n$ Kramers pairs
$\{\psi_i,\psi_{\bar i}=\mathcal{K}\psi_i\}_{i=1}^{n}$, where $\mathcal{K}$ ($=-i\mathbf{I}_2\otimes\sigma_y K_0$) is the time reversal operator
(NB:  $K_0$ represents complex conjugation).
%Due to the Hermitian nature of the Hamiltonian, the one-body term is Hermitian and the two-body term admits a 4-fold permutation symmetry:
%\begin{equation}
%	(ij|kl) = (kl|ij) = (ji|lk)^\ast = (lk|ji)^\ast
%	\label{sym_int}
%\end{equation}
%In the nonrelativistic case, spin is a good quantum number,
%the Hamiltonian can be further simplified.
%For example, since the orbitals are real, the two-electron integrals possess an additional permutation symmetry:
%\begin{equation}
%	(ij|kl)=(ij|lk)
%\end{equation}
%resulting in an overall 8-fold permutation symmetry.
%For the relativistic Hamiltonians, spin is no longer a good quantum and the time-reversal symmetry should be explored instead.
%In this case, $i,j,k,l$ refer to spinors, which can be labeled as barred or unbarred to emphasize Kramers pairs.
%By fully exploring the permutation symmetry of the two-body terms under time-reversal symmetry,
%a 16-fold reduction in storage requirements for integrals can be achieved\cite{RQC_BOOK_Dyall,4C-iCIPT2}.
%When relativistic effects are moderate, a more economical treatment of spin-orbit coupling is provided by the sf-X2C + so-DKH1 Hamiltonian\cite{X2CSOC1,X2CSOC2}, where relativistic effects are treated perturbatively.
%Crucially, for all these Hamiltonian formulations,
%the generic expression (Eq. \eqref{sym_op}/\eqref{sym_int}) hold.
%This universality forms the basis for our unified treatment of various Hamiltonians.
By making only use of the permutation symmetry of the molecular integrals, Eq. \eqref{Hamiltonian} can be decomposed as\cite{iCIPT2}
\begin{align}\label{decompositionH}
		 H&=H_1^0+H_2^0+H_1^1+H_2^1+H_2^2 \\
		 H_1^0&=\sum_i h_{i i} E_{i i} \label{H1-0}\\
		 H_1^1&=\sum_{i<j} h_{i j} E_{i j}+\sum_{i>j} h_{i j} E_{i j} \label{H1-1}\\
		 H_2^0&=\frac{1}{2} \sum_i(i i | i i) E_{i i}\left(E_{i i}-1\right)\nonumber \\
		& +\sum_{i<j}\left[(i i | j j) E_{i i} E_{j j}+(i j | j j) e_{i j, j i}\right] \label{H2-0a}\\
		 H_2^1&=\sum_{i \neq j}\left[(i i | i j)\left(E_{i i}-1\right) E_{i j}+(i j | j j) E_{i j}\left(E_{j j}-1\right)\right]\nonumber \\
		& +\sum_{i \neq j \neq k}\left[(i j | k k) E_{i j} E_{k k}+(i k | k j) e_{i k, k j}\right] \label{H2-1a}\\
		 H_2^2&=\sum_{i \leq k} \sum_{j \leq l}^{\prime}\left[2^{-\delta_{ik} \delta_{jl}}(i j | k l) e_{i j, k l}+\left(1-\delta_{i k}\right)\left(1-\delta_{j l}\right)(i l | k j) e_{i l, k j}\right]\label{H2-2}
\end{align}
where the superscripts of $H_i$ $(i=1,2)$ classify the
excitation levels (zero, single or double) between a pair of orbital configurations (oCFG),
whereas the prime in the summation of $H_2^2$ indicates that $\{j, l\} \cap\{i, k\}=\varnothing$.
The individual terms of $H_i^j$ ($i=1,2; j=0,1,2$) can be represented pictorially by the diagrams shown in Figs. \ref{Diagrams-0} to \ref{Diagrams-2},
which are drawn with the following conventions:
(1) The enumeration of orbital/spinor levels starts with zero and increases from bottom to top.
(2) The left and right vertices (represented by filled diamonds)
indicate creation and annihilation operators, respectively, which form a single generator when connected by a non-vertical line.
(3) Products of single generators should always be understood as normal ordered.
For instance, Fig. \ref{Diagrams-1}(m) means $\{E_{ik}E_{kj}\}=\{E_{kj}E_{ik}\}=E_{ik}E_{kj}-E_{ij}=e_{ik, kj}$ ($i<j<k$),
which is the exchange counterpart
of the direct generator $\{E_{ij}E_{kk} \}=\{E_{kk}E_{ij}\}=e_{ij,kk}$ ($i<j<k$) shown in Fig. \ref{Diagrams-1}(g).
The same diagrams can be used to compute the basic coupling coefficients (BCC) between spin-free\cite{Paldus1980} or
between spin-dependent\cite{4C-iCIPT2} CSFs/DETs. It is just that, in the spin-dependent case,
every vertex of the diagrams [i.e., every term in the algebraic expressions \eqref{H1-0} to \eqref{H2-2}]
is composed of both unbarred (A) and barred (B) spinors (e.g., the first term of $H_2^0$ \eqref{H2-0a} is composed of
16 terms, in stead of merely one term as in the spin-free case).
As a matter of fact, due to the underlying hermiticity, only the $s_2$, $c_x$ ($x\in[1,6]$), and $d_x$ ($x\in[1,7]$)
types of diagrams are needed for computing the one- and two-particle BCCs.
That is, once the BCCs for these types of diagrams are available, those
for the conjugate diagrams can be obtained simply by matrix conjugate transpose.
In particular, the tabulated unitary group approach (TUGA)\cite{iCIPT2,iCIPT2New} can be employed to evaluate and reuse
the BCCs between randomly selected spin-free or spin-dependent CSFs/DETs.
The decomposed form \eqref{decompositionH} of the Hamiltonian \eqref{Hamiltonian} offers an additional advantage:
by using oCFGs as the organization unit, the same set of integrals can be utilized for multiple HMEs (vide post).
In particular, some upper bounds $\tilde{H}^{IJ}$ for the HMEs
between the CSFs/DETs of oCFG pairs $|I\rangle$ and $|J\rangle$ can be established\cite{iCIPT2New} in advance (cf. Table \ref{decomposition_of_Hamiltonian}) and then
used to facilitate the selection procedure.
%Finally, it is worth noting that diagrams employed in UGA provide a unified perspective for constructing matrix elements via either Slater-Condon rules or UGA, as well as for formulating matrix product operators in the context of DMRG.
%This approach integrates these two categories of wavefunction ansätze into a unified framework.
%Further details on this aspect will be elaborated in a forthcoming paper.

\begin{figure}
	\includegraphics[width=1.0\textwidth]{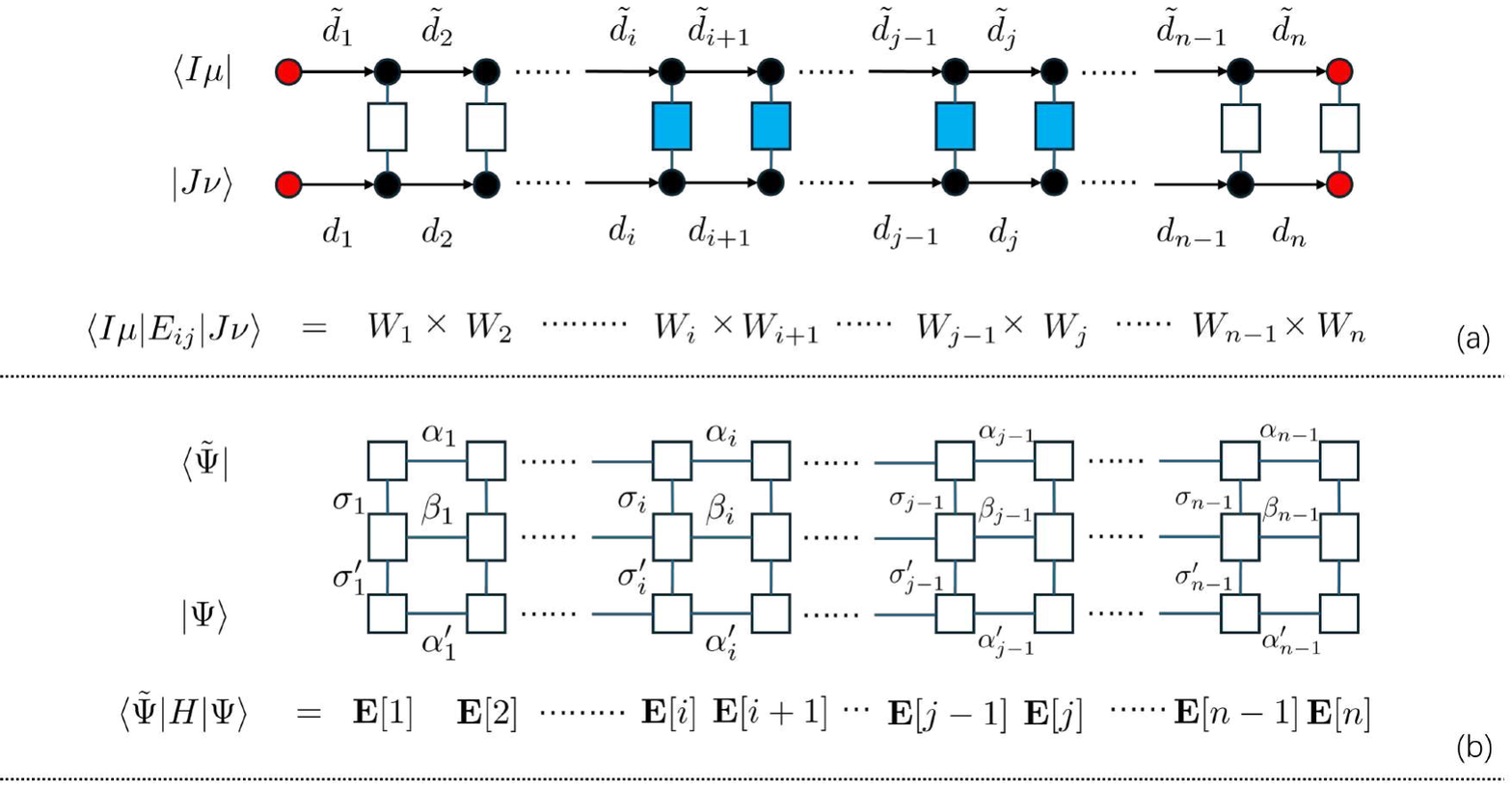}
	\caption{Graphical representations of (a) product of segment values $W_r$ ($r\in[1,n]$) in Eq. \eqref{eij_ccf} and (b) product of transfer matrices
$\mathbf{E}(i)$ ($i\in[1,n]$) in Eq. \eqref{Hexpect}. }
	\label{Similarity}
\end{figure}

It deserves to be mentioned that the above diagrammatic representation of the Hamiltonians can also be used
in DMRG working with matrix product states/operators (MPS/MPO).
To see this, recall first that the spin-free BCC $\langle I\mu|E_{ij}|J\nu\rangle$ is evaluated in UGA\cite{Paldus1980} via
product of the segment values $W_r$,
\begin{equation}
	\langle I \mu| E_{i j}|J \nu\rangle =\prod_{r=1}^n W\left(Q_r ; \tilde{d}_r d_r, \Delta b_r, b_r\right)=:\prod_{r=1}^n W_r\label{eij_ccf}
\end{equation}
where $\{d_r\}_{r=1}^n$ are the Shavitt step numbers\cite{Shavitt1977} [NB: $b_r$ can be
derived from $d_p$ ($p\in[1,r]$) and $\Delta b_r=b_r-\tilde{b}_r$), in terms of which the bar and ket CSFs can be expressed as
$|I\mu\rangle = |\tilde{d}_1\cdots\tilde{d}_n\rangle$ and
$|J\nu\rangle=|d_1\cdots d_n\rangle$, respectively. In contrast, in
the MPO/MPS formalism\cite{DMRG_MPO} of DMRG, the Hamiltonian and wavefunctions are expressed as
\begin{equation}
	H=\sum_{\{\sigma_k,\sigma_k^\prime\}}
	\left(
	\sum_{\{\beta_k\}}
	O_{\beta_1}^{\sigma_1 \sigma_1^{\prime}}[1] O_{\beta_1,\beta_2}^{\sigma_2 \sigma_2^{\prime}}[2]
	\cdots
	O_{\beta_{n-1}}^{\sigma_n \sigma_n^{\prime}}[n]
	\right)
	\left|\sigma_1\sigma_2 \cdots \sigma_n\right\rangle\left\langle \sigma_1^{\prime} \sigma_2^{\prime} \cdots \sigma_n^{\prime}\right| \label{HMPO}
\end{equation}
\begin{equation}
	|\tilde{\Psi}\rangle = \sum_{\{\sigma_k\}}
	\left(
	\sum_{\{\alpha_k\}}
	\tilde{A}_{\alpha_1}^{\sigma_1}[1]
	\tilde{A}_{\alpha_1 \alpha_2}^{\sigma_2}[2] \cdots
	\tilde{A}_{\alpha_{n-1}}^{\sigma_n}[n]
	\right)
	|\sigma_1\sigma_2\cdots\sigma_n\rangle
\end{equation}
\begin{equation}
	|\Psi\rangle =
	\sum_{\{\sigma_k\}}
	\left(\sum_{\{\alpha_k\}} A_{\alpha_1}^{\sigma_1}[1] A_{\alpha_1 \alpha_2}^{\sigma_2}[2] \cdots A_{\alpha_{n-1}}^{\sigma_n}[n]\right)
	|\sigma_1\sigma_2\cdots\sigma_n\rangle
\end{equation}
Here, $\{\sigma_k\}$ are physical bonds, whereas $\{\alpha_k\}$ and $\{\beta_k\}$ are virtual bonds.
The transition matrix element $\langle \tilde{\Psi}|H|\Psi\rangle$ can then be calculated as
\begin{equation}
\langle \tilde{\Psi}|H|\Psi\rangle=\mathbf{E}[1]\mathbf{E}[2]\cdots \mathbf{E}[n]\label{Hexpect}
\end{equation}
where the transfer matrices are defined as
\begin{equation}
	E[i]_{\alpha_{i-1}\beta_{i-1}\alpha_{i-1}^\prime,\alpha_{i}\beta_{i}\alpha_{i}^\prime }=
	\sum_{\sigma_i,\sigma_i^\prime}
	\tilde{A}^{\sigma_{i}*}_{\alpha_{i-1}\alpha_i}
	O_{\beta_{i-1}\beta_i}^{\sigma_i\sigma_i^\prime}A_{\alpha_{i-1}^\prime \alpha_{i}^\prime }^{\sigma_i^\prime}\label{Etransfer}
\end{equation}
It follows that Eq. \eqref{Hexpect} is an analog of Eq. \eqref{eij_ccf}
(see Fig.~\ref{Similarity} for their graphical representations). For instance,
if both $\tilde{\Psi}$ and $\Psi$ are single DETs and $H=E_{ij}$, then $\mathbf{E}(i)$ (will all bond dimensions equal to one)
plays the same role as the segment value $W_i$ in Eq. \eqref{eij_ccf}.
In general, the MPO representation \eqref{HMPO} of a full Hamiltonian $H$ involves splittings of
the diagrams in Figs.~\ref{Diagrams-0} to \ref{Diagrams-2} into products of partial diagrams,
each of which represents a local operator (i.e., $O_{\beta_{i-1}\beta_i}^{\sigma_i\sigma_i^\prime}$ with specific values for
the four bonds).
The direct sum of such local operators for all relevant diagrams then gives rise to the tensor operator $O[i]$.

In short, the present diagrammatic representation of the spin-free/dependent Hamiltonians
(which can be extended to include multicomponent Hamiltonians\cite{MulticomponentHBCI,MulticomponentHBCI2} and those involving electron-phonon coupling and matter-light interactions\cite{PolaritonCC,PolaritonDMRG,PolaritonResponse})
unifies the evaluation of matrix elements via Slater-Condon rules, UGA, and MPO/MPS.
Nevertheless, the formulation of spin-adapted MPOs/MPSs in terms of the Shavitt graphs\cite{Shavitt1977} remains to be further investigated.
Moreover, it should be mentioned that some tensor contraction engine should further
be incorporated into \texttt{MetaWave} for unified implementation of relativistic and nonrelativistic CC methods.

\begin{table}[htbp]
	\centering
	\caption{Correspondence between UGA Diagrams\cite{Paldus1980} (see Figs.~\ref{Diagrams-0} to \ref{Diagrams-2})  and integral categories (IntC) as well as Upper Bounds $\tilde{H}^{IJ}$ for Hamiltonian Matrix Elements over CSFs/DETs}
	\begin{tabular}{cccccc}\toprule
		\multirow{2}[0]{*}{IntC} & \multirow{2}[0]{*}{operator} & \multirow{2}[0]{*}{range} & \multirow{2}[0]{*}{diagram} & \multicolumn{2}{c}{$\tilde{H}^{IJ}$} \\\cline{5-6}
		&       &       &       & CSF   & SD \\\toprule
		\multirow{4}[0]{*}{0} & $E_{ij}$                        & $i>j$   & s2    &       &  \\
		                      & \multirow{3}[0]{*}{$e_{ik,kj}$} & $i>j>k$ & c4    &       &  \\
		                      &                                 & $k>i>j$ & c6    &       &  \\
		                      &                                 & $i>k>j$ & d2    &       &  \\\midrule
		\multirow{2}[0]{*}{1} & $e_{ij,kl}$ & \multirow{2}[0]{*}{$k>l>j>i$} & c1    & \multirow{6}[0]{*}{$\begin{array}{@{}l@{}}
				\max(|(ij|kl)+(il|kj)|, \\
				\sqrt{3}|(ij|kl)-(il|kj)|)
			\end{array}$} & \multirow{6}[0]{*}{$\max(|(ij|kl)|,|(il|kj)|)$} \\
		& $e_{il,kj}$ &       & c3    &       &  \\
		\multirow{2}[0]{*}{2} & $e_{ij,kl}$ & \multirow{2}[0]{*}{$k>l>i>j$} & d1    &       &  \\
		& $e_{il,kj}$ &       & c5    &       &  \\
		\multirow{2}[0]{*}{3} & $e_{ij,kl}$ & \multirow{2}[0]{*}{$k>i>l>j$} & d3    &       &  \\
		& $e_{il,kj}$ &       & d5    &       &  \\\midrule
		4     & $e_{ij,kj}$ & $k>j>i$ & c2    & \multirow{2}[0]{*}{$2|(ij|kj)|$} & \multirow{2}[0]{*}{$|(ij|kj)|$} \\
		5     & $e_{ij,kj}$ & $k>i>j$ & d4    &       &  \\
		6     & $e_{il,ij}$ & $i>l>j$ & d6    & $2|(il|ij)|$ & $|(il|ij)|$ \\
		7     & $e_{ij,ij}$ & $i>j$   & d7    & $|(ij|ij)|$  & $|(ij|ij)|$ \\\bottomrule
	\end{tabular}%
	\label{decomposition_of_Hamiltonian}%
\end{table}%

\include{Diagram}

\subsection{Unified Handling of MPBFs}
To achieve a unified handling of MPBFs, the oCFGs are taken as the organization unit in \texttt{MetaWave}.
An oCFG $|I\rangle$, defined by a set of occupation numbers $\in\{0,1,2\}$ for the $n$ spatial orbitals in the spin-free case
or $n$ Kramers pairs of spinors in the spin-dependent case, can be represented by a binary string
composed of an array of 64-bit unsigned integers (2 bits for each spatial orbit/Kramers pair), viz.
\begin{equation}
	\underbrace{\underbrace{(n_{31}^1 \cdots n_0^1}_{\text {64-bit integer }}, \ldots, n_{31}^k \cdots n_0^k}_{\text {array of } 64 \text {-bit integers }})=\underbrace{\left|n_0^1 \cdots n_{31}^1 \cdots n_0^k \cdots n_{31}^k\right\rangle}_{\text {orbital configuration }}
\end{equation}
where $n_p^q=0,1,2$ are represented by $(00)_2,(01)_2$, and $(11)_2$, respectively. Here, the number of bit 1
coincides with the occupation number, thereby enabling efficient bitwise operations on oCFGs\cite{iCIPT2,iCIPT2New}.
An oCFG $|I\rangle$ can generate a number of CSFs/DETs denoted as $|I\mu\rangle$, where $\mu$ distinguishes different
spin-coupling schemes for CSFs or distributions of open-shell spins/Kramers partners for DETs.

\subsection{Unified Construction of Hamiltonian Matrix}
The HMEs $\langle I\mu |H|J\nu\rangle$ involve contractions between the integrals and BCCs, the reuse of which must be maximized in sCI methods
due to the lack of structures (natural connections).
To facilitate this, we classify the HMEs into three `matrix element categories' (MEC)
corresponding to zero-, single-, and two-electron differences between oCFGs  $|I\rangle$ and $|J\rangle$,
and meanwhile classify the two-electron integrals into seven `integral categories' (IntC$\in[1,7]$ in Table \ref{decomposition_of_Hamiltonian}).
Each MEC/IntC involves only a subset of the integrals determined by the oCFG pair $(I,J)$ rather than by the individual MPBFs.
For the cases with zero- and one-electron differences, the HMEs can be simplified by introducing a closed-shell reference oCFG $|\omega\rangle$
in both the spin-free\cite{iCIPT2} and spin-dependent\cite{4C-iCIPT2} cases.

As for the reuse of the BCCs, a key observation lies in that doubly occupied or unoccupied orbitals
common to the oCFG pair $(I,J)$ can be permuted freely with other orbitals without changing the BCCs, provided
that the Yamaguchi–Kotani phase is employed in the case of CSFs\cite{Paldus1980}.
Consequently, they can be deleted, leading to a reduced oCFG pair $(\tilde{I},\tilde{J})$ spanned by
the reduced set of orbitals $\tilde{r}\in[\tilde{1},\tilde{N}]$, with
$\tilde{N}$ being the number of common singly occupied orbitals and those orbitals with different occupations in $|\tilde{I}\rangle$ and $|\tilde{J}\rangle$.
%For example, if $I=(0,1,2,1)$, $J=(0,2,2,0)$, then $\tilde{I}=(1,1)$ with $\tilde{J}=(2,0)$ and $\tilde{N}=2$, since the first and third orbital are commonly unoccupied and doubly occupied respectively.
In this way, the BCCs $\langle I\mu|E_{ij}|J\nu\rangle$ and $\langle I\mu|e_{ij,kl}|J\nu\rangle$ can simply be calculated as
$\langle \tilde{I}\mu|E_{\tilde{i}\tilde{j}}|\tilde{J}\nu\rangle$ and $\langle \tilde{I}\mu|e_{\tilde{i}\tilde{j},\tilde{k}\tilde{l}}|\tilde{J}\nu\rangle$, respectively.
The orbital sequence and occupation pattern of $(\tilde{I},\tilde{J})$ can be encoded into a reduced occupation table (ROT)\cite{iCIPT2}
as the internal representation in \texttt{MetaWave}. Different oCFG pairs but of the same ROT
share the same BCCs. Moreover, due to Hermiticity of the BCCs, only those ROTs with $\tilde{I}\geq \tilde{J}$
are needed in practical applications, thereby achieving optimal reuse of the BCCs.

The general scheme for evaluating bunches of HMEs is as follows:
\begin{enumerate}
	\item Classify the given set of oCFG pairs $\{(I,J)\}$ according to the MECs
 and then establish the ROTs for each MEC;
 \item Sort oCFG pairs $\{(I,J)\}$ according to the ROTs;
 \item Fetch the integrals according to the IntCs;
\item Evaluate the BCCs on the fly across all oCFG pairs sharing the same ROT,
and then contract with different integrals for different HMEs.
	
\end{enumerate}

In summary,  different Hamiltonians, different MPBFs, and unstructured HMEs can be treated in unified manners within \texttt{MetaWave}.
%It is worth noting that the method discussed in this section
%assumes only that the Hamiltonian is Hermitian and contains up to two-body terms.
%It is natural to extend the method to include terms containing more than two-body terms,
%% such as those Hamiltonians used in the transcorrelated method,
%but we will not explore this direction in this article.
%Furthermore, the method is independent of whether the many-electron basis functions are CSFs or DETs.
%Hence, we achieve a unification of matrix element construction that is efficient and versatile for sCIPT2 methods.

\subsection{Selection}
The aim of selection is to find iteratively an improved variational space for the expansion of the wavefunction.
It involes four key components, i.e.,
ranking, diagonalization (of $P_1^{[i]}$ in Fig. \ref{sCI_workflow}), pruning, and termination.
To simplify the notation, the variational space and its first-order interacting space (FOIS) are denoted here as $P$ and $Q$, respectively.
The former gives rise to the target states $\{E^{(0)}_k, |\Psi^{(0)}_k\rangle=\sum_{J\nu\in P}|J\nu\rangle C_{\nu,k}^J\}$.

In view of MBPT, the most rigorous ranking criterion for the importance of a CSF or DET $|I\mu\rangle\in Q$ is the absolute value of its first-order coefficient
\begin{equation}
f\left(|I \mu\rangle\right)=\max_k\left|\frac{\sum_{|J \nu\rangle\in P} H_{\mu \nu}^{I J} C^J_{\nu,k}}{E^{(0)}_k-H_{\mu \mu}^{I I}}\right|,\quad
H_{\mu \nu}^{I J}=\langle I \mu|H| J \nu\rangle \label{CIPSI_criterion}
\end{equation}
%\begin{equation}
%f\left(|I \mu\rangle\right)=\left|\frac{\sum_{|J \nu\rangle \in P} H_{\mu \nu}^{I J} C^{J}_{\nu}}{E_0-H_{\mu \mu}^{I I}}\right|,\quad
%H_{\mu \nu}^{I J}=\langle I \mu|H| J \nu\rangle \label{CIPSI_criterion}
%\end{equation}
or similarly the absolute value of its second-order energy correction to the variational energy $E_0$
\begin{equation}
\epsilon\left(|I \mu\rangle\right)=\max_k\left|\frac{|\sum_{|J \nu\rangle \in P} H_{\mu \nu}^{I J} C^{J}_{\nu,k}|^2}{E^{(0)}_k-H_{\mu \mu}^{I I}}\right|\label{CIPSI_criterionE}
\end{equation}
%\begin{equation}
%\epsilon\left(|I \mu\rangle\right)=\left|\frac{|\sum_{|J \nu\rangle \in P} H_{\mu \nu}^{I J} C^{J}_{\nu}|^2}{E_0-H_{\mu \mu}^{I I}}\right|\label{CIPSI_criterionE}
%\end{equation}
Note in passing that the perturbative estimate \eqref{CIPSI_criterionE} of $\epsilon\left(|I \mu\rangle\right)$ can also be replaced\cite{MRDCIb} with that obtained
by diagonalizing the Hamiltonian over the two-state system $\{\Psi_0, |I\mu\rangle\}$. Such `CIPSI criterion'
can be used in different ways when constructing the $P$ space iteratively. For instance, all $\{|I\mu\rangle\}\in Q$ with $f\left(|I \mu\rangle\right)$ larger than a preset threshold $C_{\mathrm{Q}}$
are retained in the variational space $P$ at each iteration of CIPSI (configuration interaction with perturbative selection made iteratively)\cite{CIPSIa}.
It is obvious that the final size of $P$ (denoted as $|P|$) is determined by the threshold $C_{\mathrm{Q}}$ itself.
A variant\cite{CIPSI-DMCa} of CIPSI is that those $\{|I\mu\rangle\}\in Q$ with the largest $\epsilon\left(|I \mu\rangle\right)$ are put into $P$
until the target size $|P|$ has reached. An additional constraint is invoked in
ASCI (adaptive sampling configuration interaction)\cite{ASCI2016}, where except for $|P|$,
$|Q|$ is also kept fixed throughout the iterative selections, by making use of $f\left(|I \mu\rangle\right)$.
A different strategy is adopted in ACI (adaptive configuration interaction)\cite{ACI2016}, where those $\{|I\mu\rangle\}\in Q$ of the smallest $\epsilon\left(|I \mu\rangle\right)$ are excluded as long as the sum of their $\epsilon\left(|I \mu\rangle\right)$ is smaller than the target accuracy of energy.
At variance with these implementations which feature two spaces ($P$ and $Q$), ICE (iterative configuration expansion)\cite{ICE-2}
adopts the three-space concept\cite{angeli1997multireference1}, where the $P$ space is split into two subspaces spanned by the
generators and the rest (spectators), which have weights larger and smaller than a threshold, respectively.
Only the former are allowed to access perturbers $\{|I\mu\rangle\}\in Q$, which are included in the $P$ space only when their $\epsilon\left(|I \mu\rangle\right)$
are larger than a second threshold.

\begin{figure}
	\includegraphics[width=0.7\textwidth]{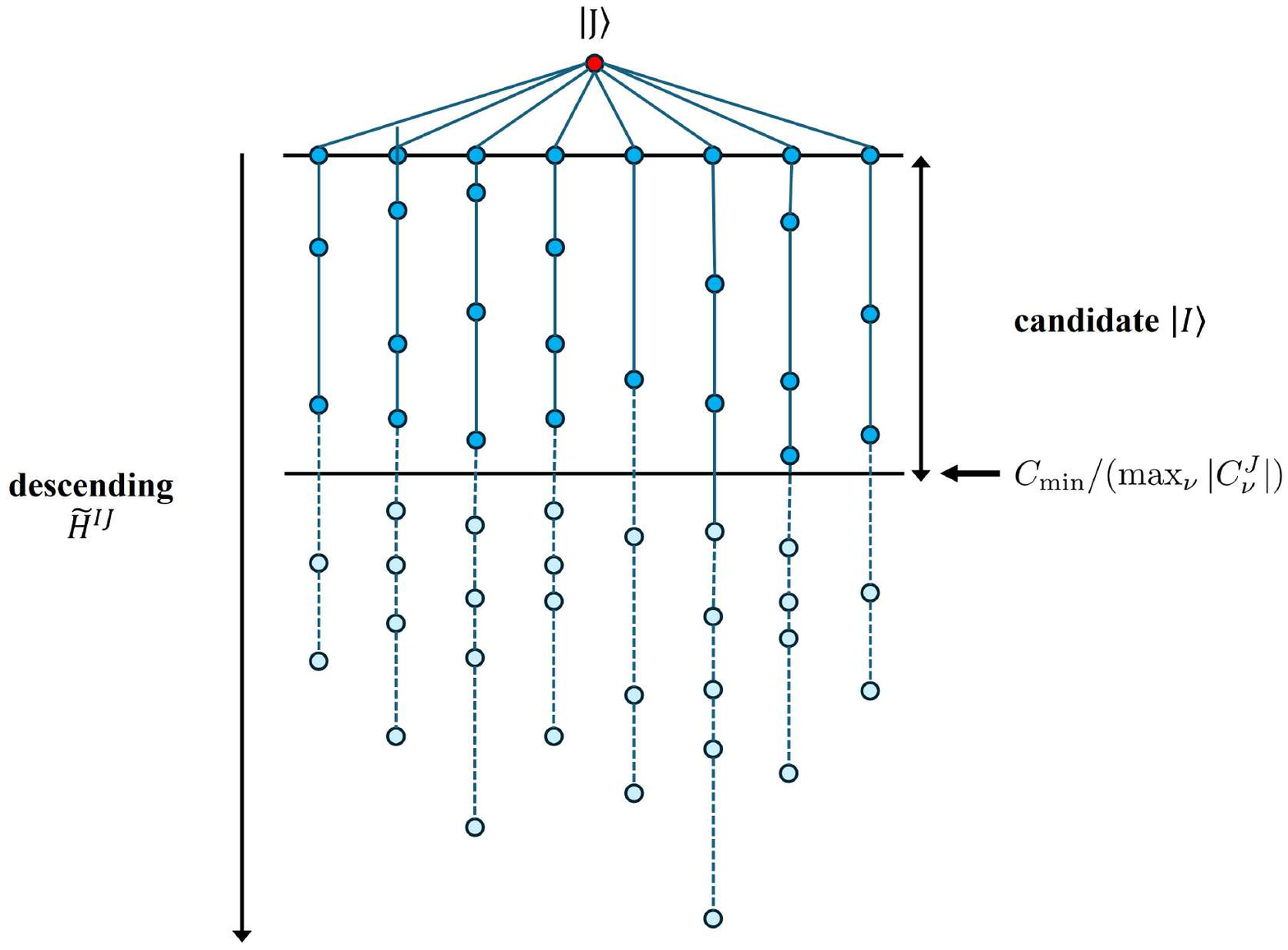}
	\caption{Screening of doubly excited configurations based on upper bounds $\tilde{H}^{IJ}$ of Hamiltonian matrix elements.
	}
\label{iCI_Selection}
\end{figure}

Noticing that the CIPSI criterion \eqref{CIPSI_criterion}/\eqref{CIPSI_criterionE} is computationally too expensive (due to the summation in
the numerator), a drastically simplified criterion was introduced in SHCI (semi-stochastic heat-bath configuration interaction)\cite{HBCI2016}, viz.
\begin{equation}
	\left.f(|I \mu\rangle,|J\rangle, \eta_Q\right)=\max_{\nu,k}\left|H_{\mu \nu}^{I J} C^{J}_{\nu,k}\right| \geq \eta_Q \label{SHCI_criterion}
\end{equation}
It follows that those $|I\mu\rangle\in Q$ with $|H^{IJ}_{\mu\nu}|<\eta_Q/\max_k|C^J_{\nu,k}|$ are never touched (cf. Fig. \ref{iCI_Selection}), provided that
the integrals are pre-sorted in descending order of their magnitudes.
However, such integral-driven selection usually leads to a variational space $P$ that is much less compact\cite{iCIPT2New}
than that by using the coefficient/energy-driven selection \eqref{CIPSI_criterion}/\eqref{CIPSI_criterionE}.
To remedy this but retaining the efficiency, the good of the integral- and coefficient-driven
selections is combined in iCIPT2\cite{iCIPT2,iCIPT2New}, by means of the following boolean function $f\left(|I \mu\rangle,|J\rangle, C_{\min }\right)$:
\begin{enumerate}[(A)]
\item If $|I\rangle$ is identical with or singly excited from $|J\rangle\in P$, then
\begin{equation}\label{iCI_Selection-1}
\begin{aligned}
	& f\left(|I \mu\rangle,|J\rangle, C_{\min }\right)=\left(\max_{\nu,k}\left(\left|H_{\mu \nu}^{I J} C^{J}_{\nu,k}\right|\right) \geq C_{\min }\right) \quad \text { and } \\
	& \left(\max_{\nu,k}\left(\left|\frac{H_{\mu \nu}^{I J} C^{J}_{\nu,k}}{E^{(0)}_k-H_{\mu \mu}^{I I}}\right|\right) \geq C_{\min }\right)
\end{aligned}
\end{equation}
\item  If $|I\rangle$ is doubly excited from $|J\rangle\in P$, then
\begin{equation}\label{iCI_Selection-2}
\begin{aligned}
	& \left.f(| I \mu\rangle,|J\rangle, C_{\min }\right)=\left(\max_{\nu,k}\left(\left|\tilde{H}^{I J} C^{J}_{\nu,k}\right|\right) \geq C_{\min }\right) \quad \text { and } \\
	& \quad\left(\max_{\nu,k}\left(\left|H_{\mu \nu}^{I J} C^{J}_{\nu,k}\right|\right) \geq C_{\min }\right) \quad \text { and } \\
	& \quad\left(\max_{\nu,k}\left(\left|\frac{H_{\mu \nu}^{I J} C^{J}_{\nu,k}}{E^{(0)}_k-H_{\mu \mu}^{I I}}\right|\right) \geq C_{\min }\right)
\end{aligned}
\end{equation}
\end{enumerate}
Literally, for case (A), loop over $|I \mu\rangle$ in $Q$ and evaluate $H_{\mu \nu}^{I J}$ for all CSFs $|J \nu\rangle \in P$.
If $\max_{\nu,k} \left(\left|H_{\mu \nu}^{I J} C^{J}_{\nu,k}\right|\right)$ is larger than $C_{\min }$ then evaluate $H_{\mu \mu}^{I I}$; otherwise discard $|I \mu\rangle$.
If $\max_{\nu,k}\left(\left|\frac{H_{\mu \nu}^{IJ} C^{J}_{\nu,k}}{E^{(0)}_k-H_{\mu \mu}^{I I}}\right|\right)$ is larger than $C_{\min }$ then $|I \mu\rangle$ is selected.
As for case (B), only those doubly excited oCFGs $|I\rangle$ with estimated upper bounds
 $\tilde{H}^{I J}$ (cf. Table \ref{decomposition_of_Hamiltonian}) larger than $C_{\min } /\max_{\nu,k} $ $|C^{J}_{\nu,k}|$ need to be generated (i.e., those unimportant ones are never touched, see Fig. \ref{iCI_Selection}). For such $\{|I\rangle\}$, the remaining step is the same as case (A).

It should be noted that not all CSFs/DETs have appreciable weights in the wavefunction ($\Psi_1^{[i]}$ in Fig. \ref{sCI_workflow}), even with the above ranking. Therefore,
those CSF/DETs of very small coefficients should be deleted to end up with
a compact varitional space ($P_2^{[i]}$ in Fig. \ref{sCI_workflow}) that is affordable for subsequent PT2 correction. Such a pruning step is invoked in ASCI
(by retaining only a fixed number of DETs of the largest coefficients)\cite{ASCI2016}, ACI (by enforcing that
the sum of the squared coefficients of DETs is smaller than a preset threshold)\cite{ACI2016}, and iCIPT2
(by retaining all CSFs with coefficients larger than $C_{\mathrm{min}}$ in absolute value)\cite{iCIPT2}, but not
in CIPSI\cite{CIPSIa,CIPSI-DMCa}, SHCI\cite{HBCI2016}, and ICE\cite{ICE-2}. While this is not very serious for CIPSI and ICE,
it is not the case for SHCI due to the use of the very loose criterion \eqref{SHCI_criterion}. It has been shown\cite{iCIPT2New} that
typically 90\% of the DETs selected according to the criterion \eqref{SHCI_criterion} can be pruned away, meaning that
a substantial time is wasted in the diagonalization. Speaking of the overall efficiency,
the whole selection procedure with the iCI criterion \eqref{iCI_Selection-1}/\eqref{iCI_Selection-2}  is about five times cheaper,
 whereas that with the CIPSI \eqref{CIPSI_criterion} or SHCI \eqref{SHCI_criterion} criterion is about two times more expensive than the PT2 correction step
 (for more details, see Ref. \citenum{iCIPT2New}).

The iterative selection procedure has to be terminated in one way or another. Unlike the energy convergence
criterion ($|E_1^{[i+1]}-E_1^{[i]}|\leq \varepsilon$) adopted in the aforementioned sCI methods, the similarity between
the start ($P_0^{[i]}$) and pruned ($P_2^{[i]}$) spaces of iteration $i$ (see Fig. \ref{sCI_workflow})
is taken as the convergence check in iCIPT2, viz.
\begin{equation}
	\frac{\left|P_0\left[C_{\text{min}}\right] \cap P_2\left[C_{\text{min}}\right]\right|}{\left|P_0\left[C_{\text{min}}\right] \cup P_2\left[C_{\text{min}}\right]\right|} \geq \tau_{\mathrm{p}}
\end{equation}
where $\tau_{\mathrm{p}}$ (=0.95) is a universal parameter.

Finally, some words on the diagonalization are necessary. As well known, the best way to make full use of
the sparsity of the Hamiltonian matrix $\mathbf{H}$ is through the matrix-vector products, $\bm{\sigma}_i=\mathbf{H}\mathbf{X}_i$,
which require only nonzero elements of $\mathbf{H}$ (which are never stored but are recalculated whenever needed). The crucial point
here is how to choose the trial (column) vectors $\{\mathbf{X}_i\}_{i=1}^{N_b}$, so as to expand the target roots
accurately even if the space $\mathbf{X}=\mathrm{span}\{\mathbf{X}_i\}$ is very incomplete.
Unfortunately, most schemes work with a search space $\mathbf{X}$ whose dimension $N_b$ increases along the iterations, thereby
leading to large memory consumptions. To avoid this, iCIPT2 adopts the iterative vector interaction (iVI) approach\cite{iVI,iVI-TDDFT},
which has the following merits: (1) it adopts well defined trial vectors (by perturbation theory, intermediate Hamiltonian, conjugate gradient, etc.);
(2) it works with a fixed-dimensional search space, with the dimension determined automatically
by the number of target roots; (3) it can access directly interior roots without touching none of the lower roots.

\subsection{Constraint-Based ENPT2}
As long as the iteratively determined variational space $P$ is good enough, the remaining dynamic correlation can be accounted for accurately at the lowest-order perturbation theory. To be more general, the three-space concept\cite{angeli1997multireference1,ICE-2} can be adopted here.
That is, the $P$ space is split into a primary ($P_m$) and a secondary ($P_s$) subspace,
and only the former is allowed to connect the external space $Q=1-P$. The eigenpairs $\{\tilde{E}^{(0)}_k, |\tilde{\Psi}^{(0)}_k\rangle=\sum_{J\nu\in P_m}|J\nu\rangle C_{\nu,k}^J\}$
of the $P_m$ space are  to be corrected by ENPT2 in a state-specific manner, viz.
\begin{equation}
\tilde{E}_{k}^{(2)}[Q]=\sum_{|I \mu\rangle \in Q} \frac{\left|\sum_{|J \nu\rangle \in P_m} H_{\mu \nu}^{I J} C_{\nu, k}^J\right|^2}{\tilde{E}_k^{(0)}-H_{\mu \mu}^{I I}}\label{ENPT2E}
\end{equation}
Given its simplicity, the implementation of Eq. \eqref{ENPT2E} encounters four issue:
\begin{enumerate}
\item How to generate efficiently the $Q$ space, given the unstructured $P_m$ space?
\item How to minimize the memory consumption, given the extremely large size of the $Q$ space?
\item How to establish the connections between the oCFGs $|I\rangle\in Q$ and those interacting oCFGs $|J\rangle\in P_m$?
\item How to compute efficiently the matrix elements?
\end{enumerate}
To address these issues, note first that the single and double excitations from the CSFs/DETs of the $P_m$ space in general
span a space $W$ that is larger than $Q$, for some of them may belong to $P_m$ or $P_s$. That is, the net $Q$ space
can only be obtained by removing the $P$-space CSFs/DETs from $W$. However, this requires a very slow double-check process.
Fortunately, this issue can be bypassed by rewriting Eq. \eqref{ENPT2E} as\cite{ASCI2018PT2}
\begin{align}
\tilde{E}_{k}^{(2)}[Q]=&\tilde{E}_{k}^{(2)}[W]-\tilde{E}_{k}^{(2)}[P], \quad W=P \cup Q=P_m\cup P_s\cup Q  \label{PT2final}\\
		\tilde{E}_{k}^{(2)}[W]=&\sum_{|I\mu\rangle \in W} \frac{\left|\sum_{|J\nu\rangle \in P_m, |J\nu\rangle \neq |I\mu\rangle} H_{\mu \nu}^{I J} C_{\nu, k}^J\right|^2}{\tilde{E}_k^{(0)}-H_{\mu \mu}^{I I}} \label{EW1}\\
		\tilde{E}_{k}^{(2)}[P]=&\sum_{|I \mu\rangle \in P}\frac{\left|\langle I \mu|H|\tilde{\Psi}_k^{(0)}\rangle
-\langle I \mu|\tilde{\Psi}_k^{(0)}\rangle H_{\mu\mu}^{II}\right|^2}{\tilde{E}_k^{(0)}-H_{\mu \mu}^{I I}} \label{EP1}
\end{align}
The negative term in the numerator of Eq. \eqref{EP1} arises from the fact that the diagonal terms
have been excluded in Eq. \eqref{EW1}. The final energies are calculated as
\begin{align}
E_k=E_k^{(0)}+\tilde{E}_{k}^{(2)}[Q]
\end{align}
where $\{E_k^{(0)}\}$ are obtained by the diagonalization of the $P$ space instead of the $P_m$ space.
If $P_m$ is just the whole $P$, Eq. \eqref{EP1} can further be rewritten as
\begin{align}
 \tilde{E}_{k}^{(2)}[P]
=&\sum_{|I \mu\rangle \in P}\left(C_{\mu, k}^I\right)^2\left(\tilde{E}_k^{(0)}-H_{\mu \mu}^{I I}\right)\label{EP2}
\end{align}
thanks to the relation $\langle I \mu| H |\tilde{\Psi}_k^{(0)}\rangle=\tilde{E}_k^{(0)}\langle I \mu|\tilde{\Psi}_k^{(0)}\rangle=\tilde{E}_k^{(0)} C_{\mu,k}^I$
 for $\forall |I\mu\rangle\in P$.

\begin{figure}
	\includegraphics[width=0.6\textwidth]{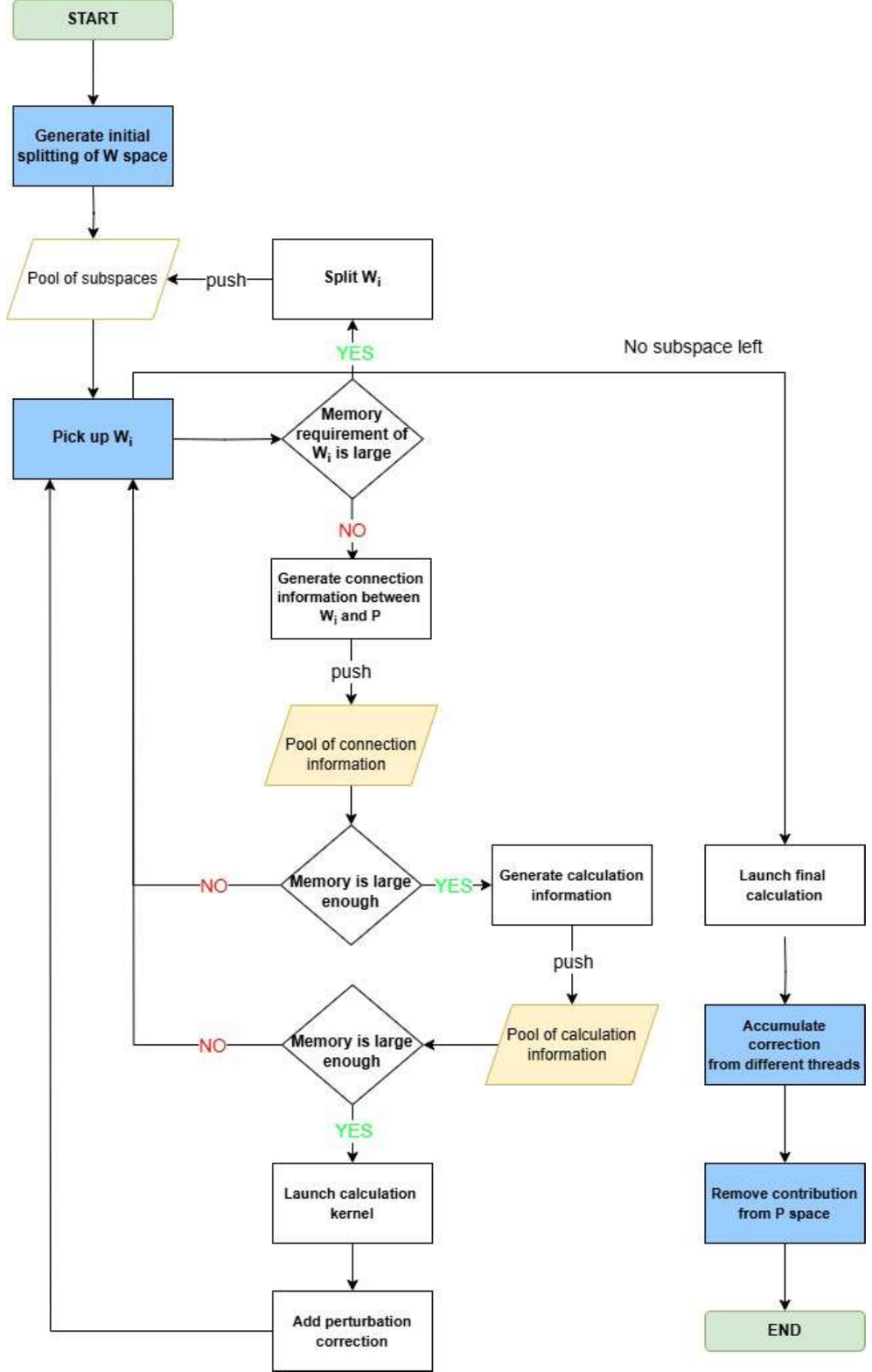}
	\caption{Workflow of pipelined, residue-based constraint-based ENPT2 in \texttt{MetaWave}.
	Green boxes represent the start and end points of the algorithm.
	Pool of subspaces store the those $W_i$ that to be generated,
	which is shared for each thread.
	Yellow boxes serves as a pool for storing connection or calculation information,
	which is private for each thread.
	Connection information refers to connected oCFG pairs and caluclation information refers to the ROT for each connected oCFG pair.
	Blue boxes indicate steps that must be executed either in single-thread mode or multi-thread mode with locks.
	% and managing memory usage across threads.
	White boxes denote steps that can be performed independently by each thread.
	}
	\label{workflowPT2}
\end{figure}

\begin{figure}
	\includegraphics[width=0.8\textwidth]{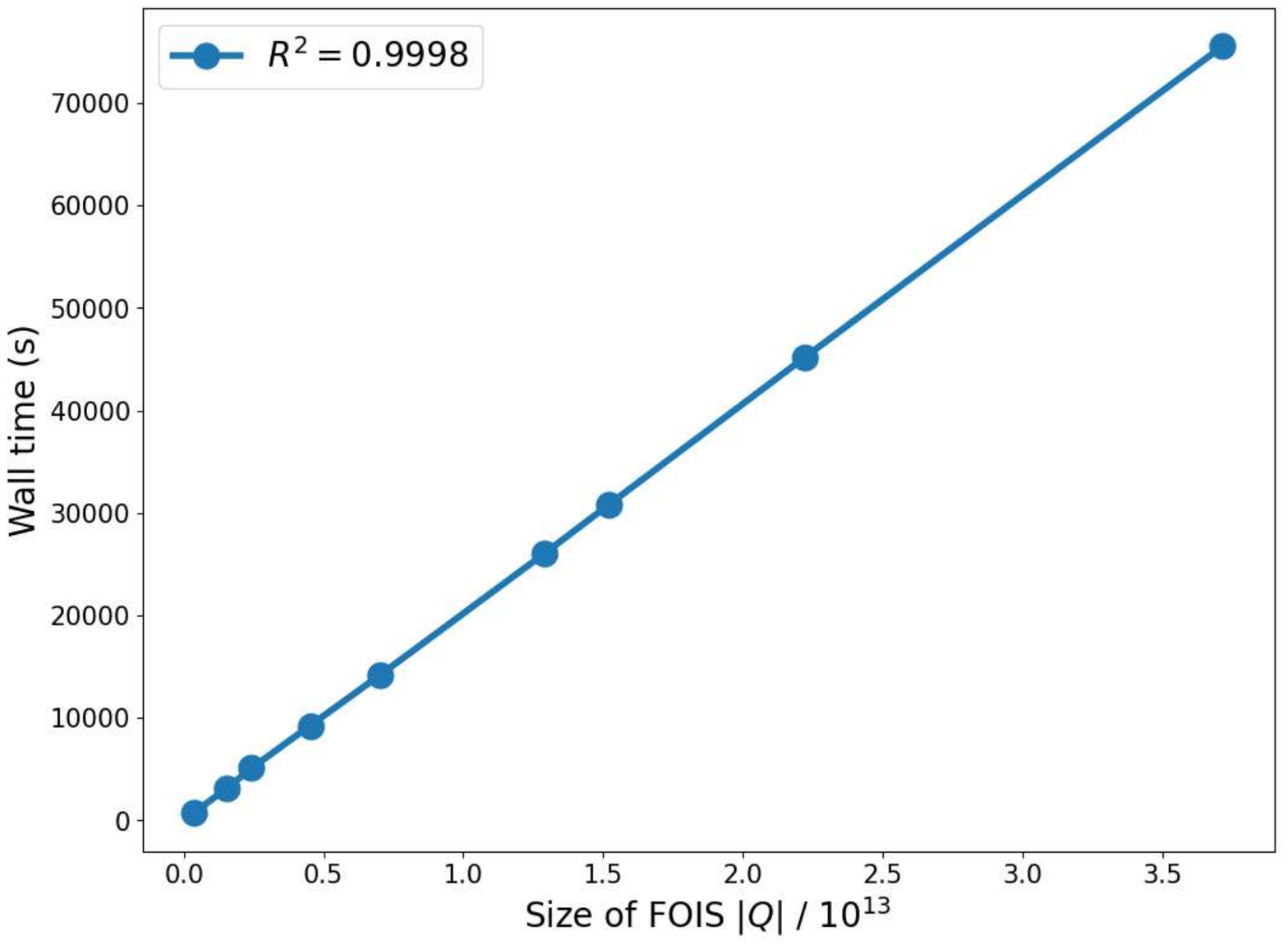}
	\caption{Wall times of the
		%\textcolor[rgb]{1.00,0.00,0.00}{PT2 step of sf-X2C-CAS(X,Y)-iCIPT2/basis}
		PT2 step of sf-X2C-CAS(28e,126o)-iCIPT2/cc-pVTZ-DK
calculations of \ce{Cr2} as a function of the size $|Q|$ of the first-order interaction space $Q$.
		The calculations were performed on a single node with two Hygon 7285 CPUs (32 cores, 2.0 GHz) and 512 GB of DDR4 memory.
	}
	\label{walltimeQ}
\end{figure}

Inspired by the idea of constraint-based PT2\cite{ASCI2018PT2},
we have developed a pipelined, constraint-based PT2 algorithm (see Fig. \ref{workflowPT2}; for more details see Ref. \citenum{iCIPT2New}):

\begin{enumerate}
	\item Split the $W$ space into disjoint subspaces $\{W_i\}$ (by constraints consisting of $L_c$ highest occupied orbitals and corresponding occupation numbers)
and dynamically distribute $\{W_i\}$ across threads or processes. Information for generating $W_i$ is stored in a pool that is shared by all threads.
	\item Picks up a subspace $W_i$ and determine the memory to store the oCFG connection information.
	%for each $W_i$.
	If $W_i$ is too large, split it into smaller subspaces until the memory requirement is met and push those smaller subspaces back to the pool and return to Step 2.
	\item Generate the oCFG connection information for each $W_i$.
	If the consumed memory does not reach the threshold, pick a new subspace and return to Step 2.
	\item
	%When the memory for storing the oCFG connection information reaches the given threshold,
	Sort the connection information to determine the oCFGs $|I\rangle$ in $W_i$ and those interacting oCFGs $|J\rangle$ in $P$. Here, those doubly excited oCFGs $|I\rangle$ with estimated upper bounds
 $\tilde{H}^{I J}$ (cf. Table \ref{decomposition_of_Hamiltonian}) smaller than $Q_{\min } /\max_{\nu,k} $ $|C^{J}_{\nu,k}|$ will be deleted,
 so as to enhance the efficiency.
	\item For each oCFG $|I\rangle$ in $W_i$, allocate memory to store the numerator and denominator
	%for the PT2 correction
	in Eq. \eqref{EW1}
	and generate the corresponding ROT for each
connected oCFG pair $(I,J)$.
	If the memory at this step reaches a predefined threshold, then launch the calculation to get the PT2 correction from all involved oCFGs.
Here, use of the sparsity of $\mathbf{H}^{IJ}$ as well as repeated use of the BCCs and molecular integrals
should be made for optimal efficiency.
	If the memory does not exceed the predefined threshold, then pick up a new subspace and go back to Step 2.
	\item After finishing $\tilde{E}_{k}^{(2)}[W]$ in Eq. \eqref{EW1}, calculate $\tilde{E}_{k}^{(2)}[P]$ in Eq. \eqref{EP1}/\eqref{EP2} and
	$\tilde{E}_{k}^{(2)}[Q]$ in Eq. \eqref{PT2final}.
\end{enumerate}
As shown in Fig. \ref{walltimeQ}, the wall time of the PT2 correction scales perfectly linearly with respect to the size of $Q$, demonstrating
the efficacy of the pipelined algorithm in memory management and massive parallelization. In particular, the algorithm is
 generic with respect to Hamiltonians and MPBFs.

\section{Software Architecture}\label{architecture}
\begin{figure}
	\includegraphics[width=0.7\textwidth]{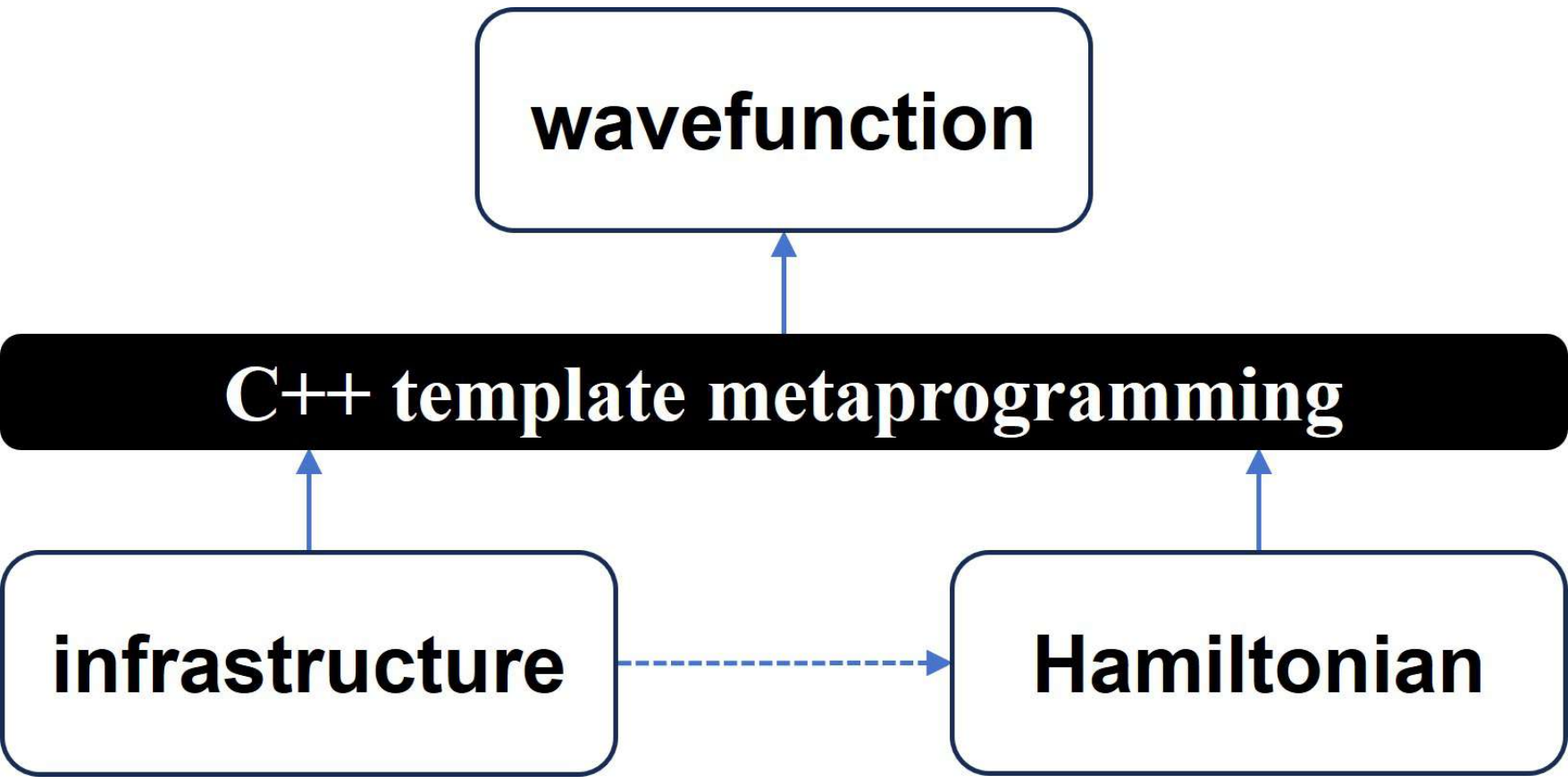}
	\caption{Three-layer architecture of \texttt{MetaWave}. 
	The  \texttt{infrastructure} layer abstracts linear algebra operations and parallelization, enabling
the  \texttt{Hamiltonian} and  \texttt{wavefunction} layers to treat scalar types and parallel strategies as template parameters.
	The  \texttt{Hamiltonian} layer encapsulates all information relevant to the Hamiltonian matrix, including molecular symmetry, MO integrals, 
MPBFs, and matrix element evaluation. 
The  \texttt{wavefunction} layer leverages
	the  \texttt{Hamiltonian} and  \texttt{infrastructure} layers  
	for unified implementation of relativistic and nonrelativistic many-electron wavefunctions.
	}
	\label{iCIPT2Arch}
\end{figure}

Redundant codes and repetitive tasks should be avoided as much as possible when developing an extendable and maintainable software package.
Therefore, the design philosophy of \texttt{MetaWave} is to maximize the use of
logically identical or similar functionalities. This can be achieved through C++ mechanisms such as polymorphism and template metaprogramming.
Specifically, a three-layer abstract architecture (i.e., \texttt{infrastructure}, \texttt{Hamiltonian}, and \texttt{wavefunction}; see Fig. \ref{iCIPT2Arch}) is employed in
\texttt{MetaWave} for furnishing a rich set of highly modular and composable functions.

\subsection{\texttt{Infrastructure} Layer}
The \texttt{infrastructure} layer serves as the basis of \texttt{MetaWave}. Designed to be independent of quantum chemistry specifics,
it provides a unified interface for basic linear algebra operations and algorithm templates for parallel computing, by
resolving the dependency of linear algebra libraries on scalar types and the complexity of managing different parallel strategies.

\subsubsection{Linear Algebra Interface}

Linear algebra operations are fundamental to computational chemistry algorithms. Optimized libraries like BLAS (Basic Linear Algebra Subprograms) are commonly used for these operations due to their efficiency. However, BLAS functions are explicitly tied to specific scalar types, such as single-precision real (\lstinline|float|), double-precision real (\lstinline|double|), single-precision complex (\lstinline|std::complex<float>|), and double-precision complex numbers (\lstinline|std::complex<double>|).
Such dependence of scalar types complicates the development of high-level codes that aim to be agnostic of the underlying data types.
Take the standard BLAS function for double-precision matrix multiplication (Listing \ref{dgemm}) as an example.
In \lstinline|cblas_dgemm| both the function name and parameters are specific to double-precision real numbers. To handle different scalar types, separate functions with different names and parameters are required, leading to code duplication and increased maintenance overhead.
To overcome this limitation, \texttt{MetaWave} introduces a unified linear algebra interface that removes the dependence of scalar types.
By utilizing C++ function overloading and templates, we provided a set of functions where the same function name can operate on different scalar types.
This allows high-level codes to perform linear algebra operations without concern for the underlying data types.

\begin{lstlisting}[caption={Declaration of cblas\_dgemm.},captionpos=t,label=dgemm]
void cblas_dgemm(const CBLAS_LAYOUT Layout,
	const CBLAS_TRANSPOSE transa,
	const CBLAS_TRANSPOSE transb,
	const MKL_INT m, const MKL_INT n, const MKL_INT k,
	const double alpha, const double *a, const MKL_INT lda,
	const double *b, const MKL_INT ldb, const double beta,
	double *c, const MKL_INT ldc);
\end{lstlisting}

\iffalse
% facilitating Hamiltonian-agnostic computations.
Moreover, we can provide a much simplified interface if the matrices are assumed to be contiguous in memory and require no transpose operations:
\begin{lstlisting}
inline void MatrixProduct(const int M, const int K, const int N, const float *A, const float *B, float *C);
inline void MatrixProduct(const int M, const int K, const int N, const double *A, const double *B, double *C);
inline void MatrixProduct(const int M, const int K, const int N, const iCI_complex_float *A, const iCI_complex_float *B, iCI_complex_float *C);
inline void MatrixProduct(const int M, const int K, const int N, const iCI_complex_double *A, const iCI_complex_double *B, iCI_complex_double *C);
\end{lstlisting}
Here, the function \lstinline|MatrixProduct| is overloaded for different scalar types.
By providing these overloaded functions, we ensure that matrix multiplication can be performed uniformly, regardless of the data type.
\fi
This unification is particularly important because nonrelativistic quantum chemical calculations typically involve real numbers, whereas relativistic calculations require complex numbers. By abstracting the scalar types, we laid the groundwork for a Hamiltonian-agnostic implementation,
allowing the same high-level code to operate seamlessly on different types of Hamiltonians.

\subsubsection{Parallel Algorithm Template}

Efficient parallel computing is crucial for the execution of wavefunction methods.
However, managing parallelism, particularly across various strategies such as OpenMP and MPI, can be complex and error-prone.
Explicitly addressing these details in each algorithm module results in code duplication and heightens the risk of bugs.
To tackle this challenge, \texttt{MetaWave} utilizes algorithm templates that encapsulate the parallelization logic.

For OpenMP, we provided a base class \lstinline|OpenMPAlgorithmBase| that defines several generic parallel execution pattern using OpenMP.
This approach allows developers to focus on the algorithmic implementation rather than the intricacies of parallelization, enhancing code
cleaning and reducing maintenance overhead. For example, in task-based parallel strategy, the core function of the generic algorithm template is the \lstinline|Run()| method
(Listing \ref{openMP}).
\begin{lstlisting}[caption={OpenMP parallel algorithm template in MetaWave.},captionpos=t,label=openMP]
inline void OpenMPAlgorithmBase::Run(){
	build_needed_info();
	omp_schedule_init();
	#pragma omp parallel num_threads(num_of_threads)
	{
		thread_init();
		#pragma omp for schedule(dynamic, 1) nowait
		for (size_t i = begin_indx; i < end_indx; ++i){
			do_task(i);}
		thread_finalize();
	}
	omp_merge_data();
}
\end{lstlisting}
By inheriting from \lstinline|OpenMPAlgorithmBase| and implementing virtual functions (e.g., \lstinline|build_needed_info()|, \lstinline|omp_schedule_init()|, \lstinline|thread_init()|, \lstinline|do_task(i)|, \lstinline|thread_finalize()|, and \lstinline|omp_merge_data()|), developers can create new algorithm modules without directly managing the parallelization details.
This design promotes code reuse and consistency across different modules, making the codebase more maintainable and less error-prone.

An additional advantage of this approach is the straightforward extension of OpenMP to MPI,
by providing an MPI-specific implementation of the base class and developing a functor that maps each OpenMP-based algorithm to its MPI-based equivalent. Specifically, two issues have to be addressed:
\begin{enumerate}
	\item How to schedule tasks on different nodes to keep load balance?
	\item How to extend shared memory algorithm to distributed memory algorithm?
\end{enumerate}
For the first issue,
it is worth noting that OpenMP algorithms rely on built-in dynamic load balancing for loops, but MPI lacks similar functionality.
To maximize parallel efficiency, we designed a double-layer dynamic task schedule framework involving both MPI and OpenMP.
%Since most MPI algorithms are based on static task distribution, typically in a ‘round-robin’ fashion, we employed dynamic task distribution at both MPI and OpenMP level.
%There is a unique MPI scheduler on the root process, and the for-loop tasks are evenly packaged into chunks, which are then dynamically dispatched by the MPI scheduler.
%The scheduling of these chunks is determined at runtime: the MPI scheduler communicates with worker nodes, tracks their statuses, and sends the next chunk to the first node that finishes its previous chunk calculation.
%Upon reaching the target node, each chunk is unpacked and processed by the original OpenMP algorithm.
%After the current chunk is finished, the worker sends a finish signal back to the root MPI scheduler, gets its next chunk, and initiates the next OpenMP calculation as the next loop cycle.
%This dynamic task schedule approach offers two main benefits: (1) it achieves perfect load balancing without the need for individual algorithm-specific settings; (2) the parallel behavior of the program can make dynamic adjustment based on runtime situation, such as variations in the actual calculation time of evenly packaged chunks, differences in the hardware and states of physical nodes and more.
%Dynamic load balancing can handle these situations during runtime perfectly.
A unique MPI scheduler on the root process packages for-loop tasks into chunks and dispatches them dynamically to worker nodes.
The scheduler monitors worker statuses and assigns the next chunk to the first available node.
Each worker unpacks the chunk and processes it using the original OpenMP algorithm.
After completing the chunk, the worker signals completion and receives the next chunk.
This approach achieves efficient load balancing without algorithm-specific settings and allows the program to adjust dynamically to runtime conditions, such as variations in computation times and hardware differences.
%So the generic OpenMP dynamic schedule will break into two parts in MPI dynamic schedule.
%The generic OpenMP dynamic schedule is thus divided into two parts in the MPI dynamic schedule, (1)
%on MPI scheduler:
Thus, the generic MPI dynamic schedule is devided into two parts, the MPI scheduler (Listing \ref{mpi_schedule}) and work nodes (Listing \ref{mpi_work_nodes}).

\begin{lstlisting}[caption={Scheduler-side code for MPI dynamic scheduling in MetaWave.},captionpos=t,label=mpi_schedule]
for (size_t chunk_id = 0; chunk_id < chunk_num; chunk_id++){
	MPI_Waitany_process();	// check if any process returns
	MPI_Send_chunk(); 	// send chunk to idle process
}
MPI_Waitall_finish();	// wait for all process finish
MPI_Send_stop_signal();    // send stop signal to workers
\end{lstlisting}
%and (2) on worker nodes:
%
\begin{lstlisting}[caption={Work nodes code for MPI dynamic scheduling in MetaWave.},captionpos=t, label=mpi_work_nodes]
while(true){
	MPI_Recv()
	if (Recv_calc_signal())	// start another original OpenMP calculation
	{
		Run_OpenMP();
	}
	else if (Recv_stop_signal())	// this worker should stop
	{
		break;
	}
	else
	{
		throw std::runtime_error();
	}
}
\end{lstlisting}
We also employed techniques like preserving thread context between OpenMP cycles and double-layer parallel data merging to optimize the MPI-OpenMP calculations.
Detailed descriptions of these optimizations will be presented elsewhere.

As for the second issue, we optimized memory usage by utilizing the serialization and memory pool techniques.
MPI communication typically requires contiguous memory, but data structures in \texttt{MetaWave} are automatically generated through inheritance, template combination, and specialization, resulting in complex, non-contiguous memory layouts.
To resolve this, we adopted a serialization approach using the cereal library\cite{cereal} for serialization and deserialization.
By specifying which class members require serialization, we wrapped MPI functions into template functions that handle serialized streams.
This allows our complex data structures to be transmitted over MPI interfaces seamlessly.
Consequently, any class with specified serializable members can be used with common MPI functions like \lstinline|MPI_Recv()| and \lstinline|MPI_Bcast()| automatically.

The use of a memory pool is to control memory usage and facilitate data reuse via pre-allocating large chunks of memory. This technique is inherently compatible with MPI.
We categorized memory usage in \texttt{MetaWave} into static memory and dynamic memory managed by the memory pool.
The former is required to store basic information (e.g., configuration spaces)
 of the program, while the latter refers to intermediate data structures (e.g., ROTs and various records used in the ranking and perturbation steps).
In our MPI implementation, the static memory is not distributed; all nodes store a copy.
The intermediate data structures in the memory pool are generated from dynamically distributed chunks, which naturally form a distributed memory structure without the need for manual intervention.
This approach optimizes memory usage and enhances performance in distributed computing environments.

\begin{figure}
	\includegraphics[width=\textwidth]{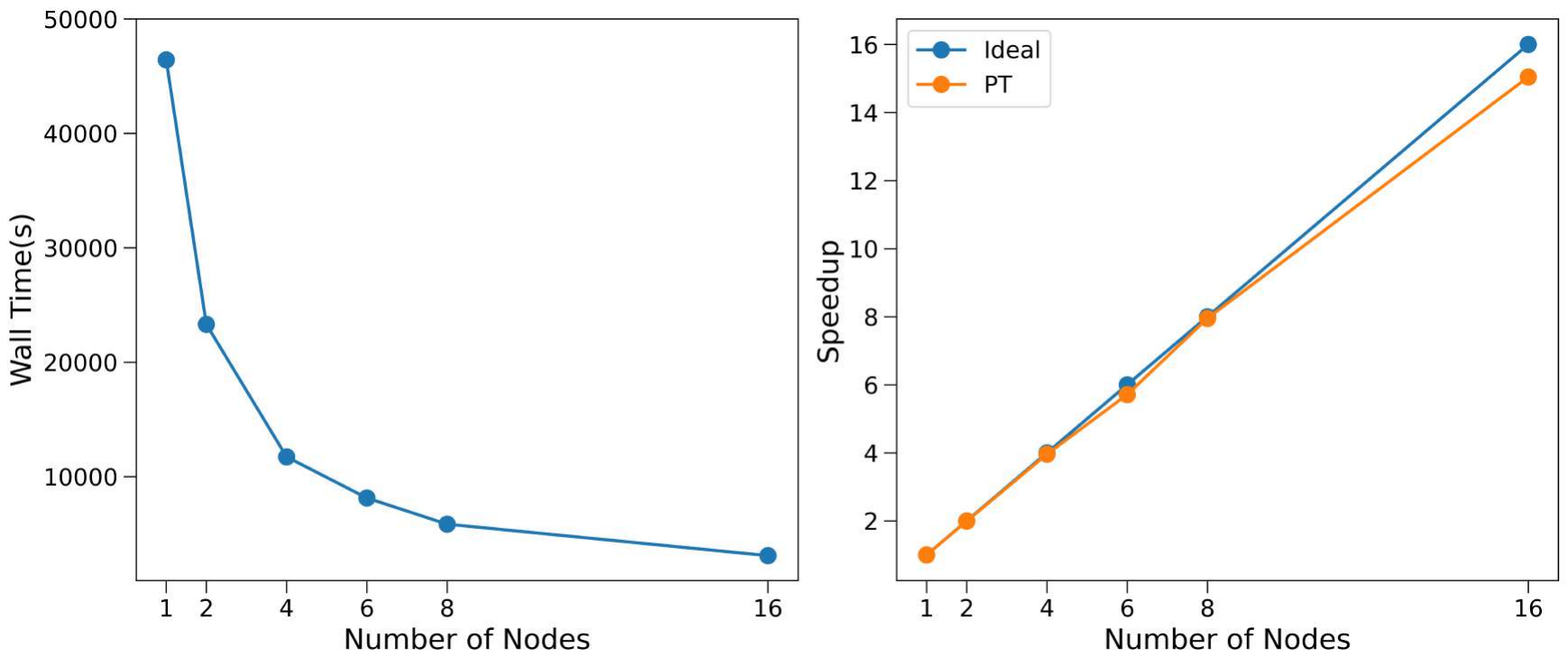}
	\caption{Wall times (left) and speedups (right) for the  PT2
	step [$E^{(2)}= -0.088495$ ${E_h}$] of sf-X2C-CAS(28e,126o)-iCIPT2[$C_{\text{min}} = 7\times 10^{-6}$]/cc-pVTZ-DK calculations of \ce{Cr2} on up to 16 nodes.
	Each node contains two Hygon 7285 CPUS (32 cores, 2.0 GHZ) and 512 GB DDR4 memory.	The variational space contains $2.0\times 10^{7}$ CSFs.
	94.0\% parallel efficiency is achieved for 16 nodes.
	}
	\label{MPI}
\end{figure}

As a demonstration of our MPI implementation, we performed CAS(28e,126o)-iCIPT2 calculations on \ce{Cr2} using the spin-free (sf) part\cite{X2CSOC1,X2CSOC2} of the
exact two-component (X2C) Hamiltonian\cite{X2C1,X2C4} and
the cc-pVTZ-DK basis set\cite{ccpv-dk}.
There are $2.0\times 10^{7}$ CSFs in the variational space with $C_{\text{min}} = 7\times 10^{-6}$ for the control of selection.
It can be seen from Fig. \ref{MPI} that our MPI implementation of the PT2 step achieved 94.0\% parallel efficiency for 16 nodes (each of which
has two Hygon 7285 CPUs (32 cores, 2.0 GHZ) and 512 GB DDR4 memory).
	
% add code snippets

\subsection{\texttt{Hamiltonian} Layer}

The \texttt{Hamiltonian} layer serves as the core of \texttt{MetaWave} and is responsible for handling domain-specific components such as Hamiltonian, MPBFs, (abelian) spatial symmetry, (non-abelian) spin symmetry, molecular integrals, and matrix element evaluation.
It creates a flexible and extensible framework that can be reused across different quantum chemical methods.
A key feature of this layer is the use of a type-trait and tagging system, which
allows for the unified handling of various Hamiltonians, molecular symmetries, and MPBFs.
By passing information through template parameters and utilizing compile-time mechanisms, both duplication reduction and maintainability enhancement
can be achieved. Furthermore, compile-time optimization is also facilitated in this way.

\subsubsection{Type-Trait and Tagging System}

To achieve flexibility and extensibility, \texttt{MetaWave} employs a type-trait and tagging system.
This system encodes properties of different Hamiltonians, molecular symmetries, and MPBFs at compile time, allowing the compiler to generate optimized code.
As a result, this type-trait and tagging system allows us to write generic code that adapts to different Hamiltonians without code duplication, in the same
spirit as other codes\cite{HORTON,PyBEST1,PyBEST2,DIRAC}.

A tag is implemented using an empty struct that serve as a unique type representing the specific concept.
For example, tags for different Hamiltonians are defined in Listing \ref{tag_H}, where \texttt{ElectronNonRela} and \texttt{ElectronRela} represent the electron-only nonrelativistic and relativistic Hamiltonians, respectively,
while \texttt{Hubbard} represents the Hubbard model Hamiltonian.
These tags are used as template parameters to differentiate between Hamiltonian types in a type-safe manner.

\begin{lstlisting}[caption={Hamiltonian tag system in MetaWave.},captionpos=t, label=tag_H]
namespace tag {
	namespace Hamiltonian {		
		struct ElectronNonRela{};
		struct ElectronRela{};
		struct Hubbard{};
	};
};
\end{lstlisting}

\begin{figure}
	\includegraphics[width=0.8\textwidth]{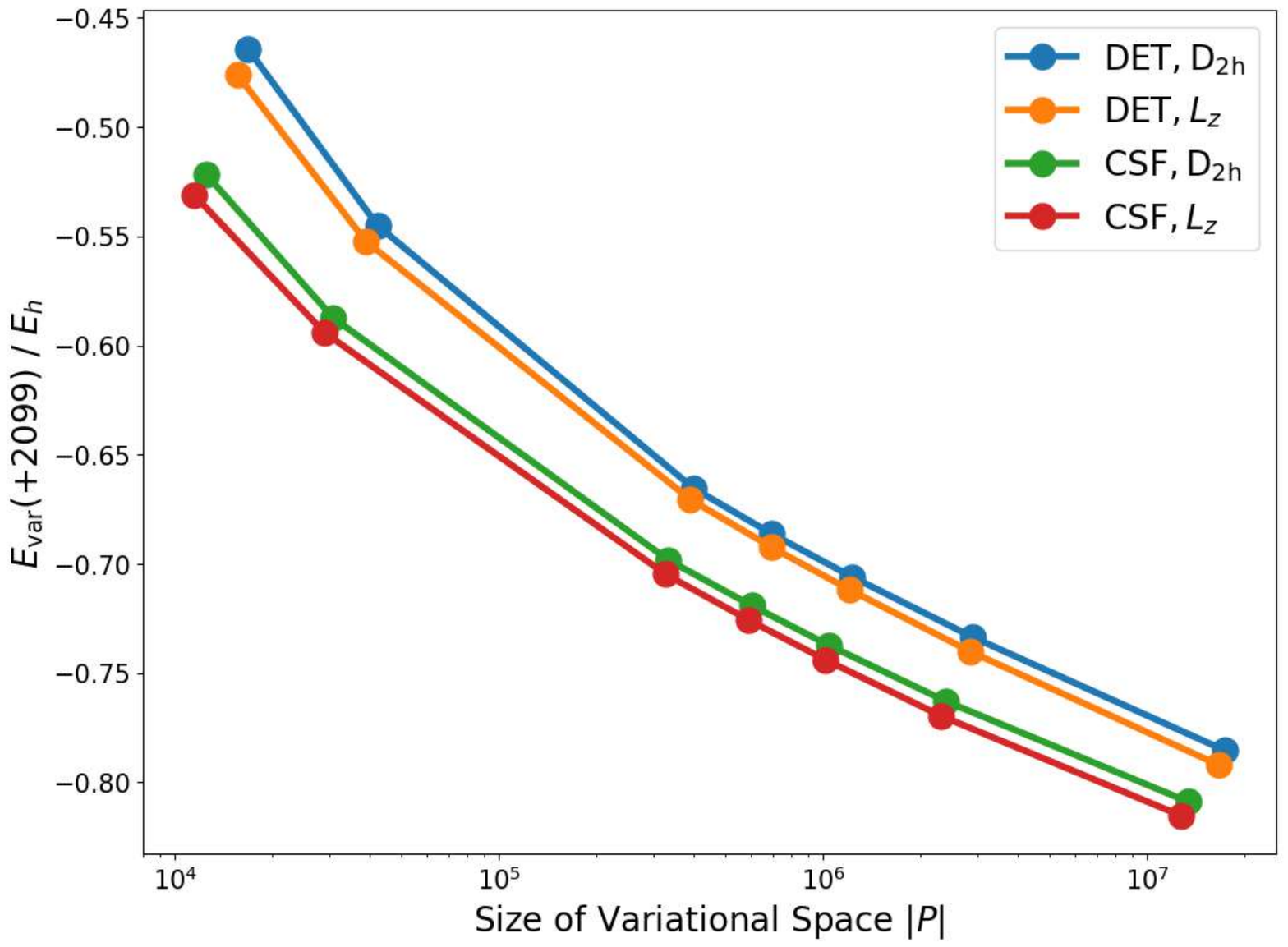}
	\caption{
	sf-X2C-CAS(28e,76o)-SiCI/cc-pVDZ-DK results for \ce{Cr2} with different symmetries.
	}
	\label{sym}
\end{figure}

To associate properties with these tags, we used type-traits structures (helper classes) that provide compile-time information about the Hamiltonian's characteristics (Listing \ref{traits_H}).
\begin{lstlisting}[caption={Hamiltonian type-traits system in MetaWave.},captionpos=t, label=traits_H]
namespace trait {
template <typename HamiltonianTy>
struct hamiltonian_property;
	
template <>
struct hamiltonian_property<tag::Hamiltonian::ElectronNonRela>
{
	static constexpr bool has_twobody_term  = true;
	static constexpr bool has_onebody_term  = true;
	static constexpr bool has_oneext_term   = true;
	static constexpr bool has_twoext_term   = true;
};

template <>
struct hamiltonian_property<tag::Hamiltonian::Hubbard>
{
	static constexpr bool has_twobody_term  = true;
	static constexpr bool has_onebody_term  = true;
	static constexpr bool has_oneext_term   = true;
	static constexpr bool has_twoext_term   = false;
};
};
\end{lstlisting}
These structures specify whether the Hamiltonian includes certain terms,
such as the one-body $H_1^0+H_1^1$ and two-body $H_2^0+H_2^1+H_2^2$ terms.
To facilitate easy access to these properties, we introduced the metafunctions\footnote{A metafunction is a function or operation that works at compile time  rather than at runtime. In C++ programming, especially in template metaprogramming, metafunctions are used to perform computations, checks, or transformations on types and constants at compile time. This allows the compiler to make decisions and optimize code before it runs, which can lead to more efficient and flexible programs.} (Listing \ref{metafunc_H}).
\begin{lstlisting}[caption={Hamiltonian metafunctions in MetaWave.},captionpos=t, label=metafunc_H]
namespace metafunc {
	
	template <typename HamiltonianTy>
	inline static constexpr bool has_twobody_term_v = trait::hamiltonian_property<Hamiltonian_t>::has_twobody_term;
	
	template <typename HamiltonianTy>
	inline static constexpr bool has_onebody_term_v = trait::hamiltonian_property<Hamiltonian_t>::has_onebody_term;
	
	template <typename HamiltonianTy>
	inline static constexpr bool has_oneext_term_v = trait::hamiltonian_property<Hamiltonian_t>::has_oneext_term;
	
	template <typename HamiltonianTy>
	inline static constexpr bool has_twoext_term_v = trait::hamiltonian_property<Hamiltonian_t>::has_twoext_term;
};
\end{lstlisting}
These compile-time constants can be used within templates to enable or disenable code paths, ensuring that only relevant computations are performed for a given Hamiltonian type.
For example, for the Hubbard model, all the code paths related to the evaluation of matrix elements between doubly excited configurations can be ignored, leading to
a more efficient implementation.
Moreover, by utilizing the type-trait and tagging system, both spatial and spin symmetries can be readily be handled.
As shown in Fig. \ref{sym}, symmetry adaptation improves significantly the compactness of the variational wavefunctions in selected iCI (SiCI) calculations.
Stated differently, on average, a symmetry adapted MPBF tends to contribute more to correlation than the unadapted one (see also Ref. \citenum{ICE-2}).

% micro level %

The type-trait and tagging system also helps to automatically generate a large number of functions with similar computational processes.
Consider, e.g., the BCCs $\langle \tilde{I}\mu|e_{\tilde{i}\tilde{j},\tilde{k}\tilde{l}}|\tilde{J}\nu\rangle$ of the reduced oCFG pair $(\tilde{I}, \tilde{J})$.
They involve segment values of all spatial orbitals, under the restriction that outside the interaction region, $[\min(\tilde{i},\tilde{j},\tilde{k},\tilde{l}), \max(\tilde{i},\tilde{j},\tilde{k},\tilde{l})]$, of the operator $e_{\tilde{i}\tilde{j},\tilde{k}\tilde{l}}$,
 the bra and ket orbitals and their occupations should be the same. The problem lies in that
there are seven different types of ROTs and various segment types (cf. Table 2 in Ref. \citenum{iCIPT2}).
It is extremely tedious and error-prone to write each case individually. This is avoided in
\texttt{MetaWave} by using template functions, where each involved component is determined automatically via the corresponding type-trait and tagging system designed for UGA.

Take the function (Listing \ref{makedbl}) for generating the BCCs between doubly excited oCFG pair $(\tilde{I}, \tilde{J})$ as an example.
Given the input parameter \lstinline|csfindx| (which corresponds to $\nu$),
it generates all non-vanishing BCCs $\langle \tilde{I}\mu|e_{\tilde{i}\tilde{j},\tilde{k}\tilde{l}}|\tilde{J}\nu\rangle$
%given $\nu$ (which corresponds to the input parameter \lstinline|csfindx|)
and store the results and the indices $\mu$ in \lstinline|drct| and \lstinline|indx| respectively.
If exists, the function also generates $\langle \tilde{I}\mu|e_{\tilde{i}\tilde{l},\tilde{k}\tilde{j}}|\tilde{J}\nu\rangle$ and stores it in \lstinline|xchng|
Since the number of CSFs that one oCFG can generate cannot exceed the range of \lstinline|int32_t| in routine applications, \lstinline|int32_t| is used to represent the indices $\mu$ and $\nu$.
In \lstinline|maketable_dbl|, \lstinline|_init_propagate| and \lstinline|_final_propagate| correspond to the evaluation outside the interaction region, while the other functions handle the evaluation within each region.
Specifically, \lstinline|_propagate_segment2| evaluates the orbitals where the operators act, whereas  \lstinline|_propagate_segment1| evaluates those where no operator acts.
To determine the correct segment value,
each function takes \lstinline|TableTy| as one of the template parameters and uses the corresponding type-traits for \lstinline|TableTy|, along with internal information contained in the class \lstinline|TUGATableMaker|.
The number specified as the second template parameter distinguishes different regions for a given operator.

\begin{lstlisting}[caption={Template function for generating BCCs between doubly excited oCFG pairs.},captionpos=t,label=makedbl]
template <typename TableTy>
inline int32_t TUGATableMaker::maketable_dbl(const int32_t csfindx, double *drct, double *xchng, int32_t *indx)
{
	if (_init_propagate<TableTy>(csfindx) == false)
	{
		return 0;
	}
		
	_propagate_segment2<TableTy, 0>();
	_propagate_segment1<TableTy, 1>();
	_propagate_segment2<TableTy, 1>();
	_propagate_segment1<TableTy, 2>();
	_propagate_segment2<TableTy, 2>();
	_propagate_segment1<TableTy, 3>();
	_propagate_segment2<TableTy, 3>();
		
	return _final_propagate<TableTy>(csfindx, drct, xchng, indx);
}
\end{lstlisting}

Labeling specific concepts via empty structs and extracting/accessing properties via type traiting/metafuncions
have been employed in all aspects of \texttt{MetaWave}, from the lowest \texttt{infrastructure} layer to the highest \texttt{wavefunction} layer.
This is the essential trick that makes each component independent and the coding as generic as possible.

\subsubsection{Configuration Space}

% 在这里着重介绍如何进行类型绑定 %

The \texttt{CfgSpace} class in \texttt{MetaWave} is a central component that manages both the MPBFs and CI vectors. It
is determined by 4 template parameters:
\begin{enumerate}
	\item \texttt{SymTy} : type of spatial symmetry;
	\item \texttt{IndxTy} : type used to store the indices $\mu$ of $|I\mu\rangle$;
	\item \texttt{BasisTy} : type of MPBFs (CSF or DET);
	\item \texttt{ScalarTy} : scalar type of CI vectors (e.g. \lstinline|double|, \lstinline|std::complex<double>|).
\end{enumerate}
The first two parameters are related to oCFGs and are hence encapsulated in the \lstinline|BasisMetadata| class (Listing \ref{BasisMetadata}) to represent the MPBFs.
This class stores the essential information of each MPBF, including spin information ($2S$ for CSFs and $2S_z$ for DETs),
number of open-shell electrons, irreducible representations (irrep), a pointer to the string that stores the oCFG, and CSF/DET indices.
%Since multiple CSFs or DETs may share the same oCFG, we use a pointer to reference the shared oCFG string, thereby reducing memory requirements.

\begin{lstlisting}[caption={Definition of class BasisMetadata representing the MPBFs.},captionpos=t,label=BasisMetadata]
template <typename SymTy, typename IndxTy>
class BasisMetadata
{
	public:
	/* typedef */
	using indx_t = IndxTy;
	using sym_t  = SymTy;
	/// .... omitted
	protected:
	int8_t    spintwo_or_sztwo, nopen;
	SymTy     irrep;
	uint64_t* oCFG;
	IndxTy    indx;
};
\end{lstlisting}

The \texttt{BasisManager} class (Listing \ref{BasisManager}) manages a collection of \texttt{BasisMetadata} instances.
It provides functionalities for handling MPBFs, such as adding CSFs/DETs, extracting information from a given CSF/DET, or sorting the MPBFs to canonical form.
This class is uniform for CSFs and DETs.
\begin{lstlisting}[caption={Definition of class BasisManager.},captionpos=t,label=BasisManager]
template <typename BasisMetadataTy>
class BasisManager
{
	public:
	/* typedef */
	
	using sym_t  = typename BasisMetadataTy::sym_t;
	using indx_t = typename BasisMetadataTy::indx_t;
	using metadata_t = BasisMetadataTy;
	using iter_t = typename std::vector<BasisMetadataTy>::iterator;
	
	/* functions */
	void add_basis(std::vector<int8_t>& Occ, uint32_t norb, uint32_t spintwo, uint32_t _indx);
	void extract_info_given_basis(const uint32_t id, uint64_t* OccCode);
	template <class CompareFuncTy>
	void sort();
	virtual void build();
	/// .... omitted
};
\end{lstlisting}

Finally, the \texttt{CfgSpace} class (Listing \ref{CfgSpace}) represents the entire configuration space, managing both the configurations and the associated CI vectors.
In \texttt{CfgSpace}, the key types, \texttt{scalar\_t}, \texttt{sym\_t}, \texttt{indx\_t}, and others, are determined by the template parameters and propagated via the \texttt{using} declarations.
This design ensures that the configuration space is aware of the types used in both the basis manager and vector manager, providing the necessary flexibility for the entire algorithm.
The use of template parameters and type binding allows for generic programming, reducing code duplication and enhancing maintainability.

\begin{lstlisting}[caption={Definition of class CfgSpace.},captionpos=t,label=CfgSpace]
template <class BasisManagerTy, class VecManagerTy, typename BasisTy>
class CfgSpace
{
	public:
		
	/* typedef */
		
	using scalar_t = typename VecManagerTy::scalar_t;
	using sym_t    = typename BasisManagerTy::sym_t;
	using indx_t   = typename BasisManagerTy::indx_t;
	using metadata_t = typename BasisManagerTy::metadata_t;
	using iter_t  =  typename BasisManagerTy::iter_t;
	using basis_t = BasisTy; // DET or CSF
		
	/// .... omitted
}:
\end{lstlisting}

\texttt{MetaWave} also provides numerous functions, such as printing for debugging purposes,
dumping data for restarting computations, merging two configuration spaces, and more.
These functionalities enhance the usability and robustness of the \texttt{MetaWave} package, allowing users to efficiently manage and manipulate the configuration space as needed.

\subsubsection{Integral Manager}

The \texttt{IntegralManager} class is responsible for handling molecular integrals.
The main challenge is how to deal with different kinds of permutation and
single/double point group symmetries.
In the spin-free case, the two-electron integrals exhibit 8-fold permutation symmetry and are real;
for systems with $L_z$ symmetry, this symmetry is reduced to 4-fold, but the integrals still remain real.
In the spin-dependent case, the integrals (over spinors) are generally complex with complicated permutation and double point group symmetries\cite{4C-iCIPT2}.
Hence, permutation and single/double point group symmetries are not uncorrelated.
To store and retrieve different kinds of molecular integrals in a uniform manner, we used a tag system to label different permutation symmetries (Listing \ref{tag_perm}) and a metafunction \lstinline|perm_symm_t| to map each single/double group irrep
to the corresponding permutation symmetry for molecular integrals at compile time.
\begin{lstlisting}[caption={Tag system to label permutation symmetries.},captionpos=t,label=tag_perm]
namespace tag {
	namespace Integrals {
		struct PermutationSymmetry
		{
			struct Fold8 {};
			struct Fold4 {};
			/// .... omitted
		}
	}
};
\end{lstlisting}

The skeleton code for the \texttt{IntegralManager} class is shown in Listing \ref{IntegralManager}.
The template parameters \lstinline|ScalarTy| and \lstinline|perm_symm_t<SymTy>| of
\lstinline|VectorStorage| specify scalar type (real or complex) and permutation symmetry of the integrals, respectively.
Note in passing that point group symmetries are not employed here for storage reduction, simply because
all integrals can be stored in memory after accounting for permutation symmetries (for the number of correlated orbitals is usually much less 1000).
Yet, point group symmetries are indeed ensured by the \lstinline|get_1b| and \lstinline|get_2b| functions,
which return zeros when the direct product of the irreps of the input orbitals does not contain the trivial irrep.
With this design, the C++ compiler can use the tagging and type-trait system to synthesize the correct data structures for storing the molecular integrals at compile time. This allows the programmer to write a uniform code without the need to consider the permutation symmetry explicitly.

\begin{lstlisting}[caption={Skeleton of class IntegralManager.},captionpos=t,label=IntegralManager]
template <typename ScalarTy, typename SymTy>
class IntegralManager
{	
	public:
		
	/* function */
		
	virtual ScalarTy get_1b(const unsigned p, const unsigned q) const
	{
		return integrals_1b(p,q);
	}
	virtual ScalarTy get_2b(const unsigned p, const unsigned q, const unsigned r, const unsigned s) const
	{
		return integrals_2b(p,q,r,s);
	}
		
	/// .... omitted
		
	protected:
		
	/* storage */
		
	VectorStorage<ScalarTy, perm_symm_t<SymTy>> integrals_1b;
	VectorStorage<ScalarTy, perm_symm_t<SymTy>> integrals_2b;
		
	/// .... omitted
};
\end{lstlisting}

\subsubsection{Evaluation of HMEs}

% emphasize type dispatching %

In \texttt{MetaWave}, the calculation of HMEs involves several steps.
The general workflow begins with the classification of oCFG pairs and generation of the corresponding reduced oCFG pairs.
This is followed by retrieving the necessary integrals, computing the BCCs, and finally contracting these components to obtain the HMEs.
Due to the complexity and intricacy of the algorithms involved in these steps, we do not delve into details here.
Instead, we emphasize that \texttt{MetaWave} provides kernel-level template functions for each step, which can be utilized by higher-level algorithms.
For example, for singly excited oCFG pairs, we have several functions shown in Listing \ref{SingleHME}.
This modular design allows for flexibility and reusability, making it easier to maintain and extend the codebase.

\begin{lstlisting}[caption={Functions for evaluation of HMEs between singly excited oCFG pairs.},captionpos=t,label=SingleHME]
template <typename TableTy>
inline void _generate_table_sngl(const uint64_t* BraCFG, const uint64_t* KetCFG, const int LenCode,
TableTy* Table, tag::Hamiltonian::ElectronNonRela);
	
template <typename TableTy>
inline void _generate_table_sngl(const uint64_t* BraCFG, const uint64_t* KetCFG, const int LenCode,
TableTy* Table, tag::Hamiltonian::ElectronRela);
	
template <typename TableTy, typename IntegralsTy>
inline void _fetch_ints_sngl(TableTy* Table, const IntegralsTy* Integrals, tag::Hamiltonian::ElectronNonRela);
	
template <typename TableInfoTransferTy, typename IntegralsTy>
inline void _fetch_ints_sngl(TableTy* Table, const IntegralsTy* Integrals, tag::Hamiltonian::ElectronRela);
\end{lstlisting}

An important aspect of our implementation is the accommodation of different Hamiltonians.
Different Hamiltonians correspond to distinct molecular integrals, which have different permutation
symmetries, and therefore different storage requirements, necessitating specialized implementations for each.
To handle this variety without code duplication, \texttt{MetaWave} employs a technique known as tag dispatching,
where specific types (tags) are used to direct the compiler to appropriate function implementations at compile time via function overloading.
For instance, the last parameter in the
functions \lstinline|_generate_table_sngl| and \lstinline|_fetch_ints_sngl|
is a tag that specifies the type of Hamiltonian, which
allows the compiler to pick up  automatically the right Hamiltonian.

\subsection{\texttt{Wavefunction} Layer}

The \texttt{wavefunction} layer is the highest level in the \texttt{MetaWave} architecture.
It employes the classes defined in the \texttt{Hamiltonian} layer as template parameters to generate the corresponding algorithm modules,
such as selection and perturbation (see Sec. \ref{Background}).
These algorithm classes follow a consistent pattern:
First, each algorithm is encapsulated in a class (typically ending with \lstinline|ThreadContext|) that describes the tasks to be performed by a single thread.
For instance, in the ranking step of the selection process, the \lstinline|RankingThreadContext| class is invoked to generate important
excited CSFs/DETs from a reference CSF/DET.
In the PT2 module, a similar class handles the calculation of the PT2 correction from a specific subspace $W_i$.
Second, parallelization is achieved through parallel templates that describe how to distribute and manage tasks across multiple threads or processes.
For OpenMP, these algorithm classes inherit from a base class (e.g., \lstinline|OpenMPAlgorithmBase| described above) and implement the necessary functions to assign tasks to each thread and merge the results efficiently.
An important aspect of our design is the implementation of MPI:
The same OpenMP parallel algorithm classes are inherited, with suitable modifications of the relevant functions to create an automatic mapping functor from OpenMP to MPI.
Since the sCI algorithms can be designed such that subtasks are independent of each other
and require no communication (aside from dynamic scheduling) during execution—only data merging in the end, this functor can effectively distribute tasks across MPI processes.
Separating MPI from the single-threaded algorithms ensures that the core algorithm logic remains unchanged while parallelism is handled independently.
Specific implementation of the \texttt{wavefunction} layer is skipped here,
as it would largely repeat the content already covered in Sec. \ref{Background}.

\subsection{Overview of \texttt{MetaWave} Functionality}
\texttt{MetaWave} offers a comprehensive suite of features designed to facilitate the development of quantum chemical methods.
The key functionalities currently supported by \texttt{MetaWave} include:
\begin{enumerate}
	\item \textbf{Hamiltonian}:
	\begin{itemize}
		\item Nonrelativistic Hamiltonian: iCIPT2\cite{iCIPT2,iCIPT2New}
		\begin{itemize}
			\item Spin symmetry;
			\item Spatial symmetry:
			\begin{itemize}
				\item Binary point group symmetry;
				\item $L_z$ symmetry;
				\item Translation symmetry (1D, 2D, and 3D);
			\end{itemize}
		\end{itemize}
		\item spin-separated X2C Hamiltonian\cite{X2CSOC1,X2CSOC2}: SOiCI and iCISO\cite{SOiCI}
		\begin{itemize}
			\item Time reversal symmetry;
			\item Binary double point group symmetry;
		\end{itemize}
		\item Relativistic Hamiltonians: 4C-iCIPT2\cite{4C-iCIPT2}
		\begin{itemize}
			\item Time reversal symmetry;
			\item Binary double point group symmetry;
		\end{itemize}
		\item Hubbard model:
		\begin{itemize}
			\item Nearest and next nearest hopping terms;
			\item 1D, 2D, and 3D;
			\item Open and periodic boundary conditions.
		\end{itemize}
	\end{itemize}
	\item \textbf{Density Matrices}:
	\begin{itemize}
		\item Reduced density matrices up to fourth order;
		\item Transition density matrices up to second order;
		\item Spin density matrices up to second order.
	\end{itemize}
	\item \textbf{Natural Orbitals} from spin-adapted first-order density matrix;
	\item \textbf{Core-Valence Separation} for core excited and ionized states;
	\item \textbf{Parallelization}:
	\begin{itemize}
		\item OpenMP
		\item MPI
	\end{itemize}
\end{enumerate}
%Since \texttt{MetaWave}  is fully generic and highly adaptable,  other wavefunction methods can readily be implemented.

\section{Application}\label{Application}
\texttt{MetaWave} has served as the platform for the development of spin-free\cite{iCIPT2,iCIPT2New} and spin-dependent\cite{SOiCI,4C-iCIPT2} iCIPT2,
which can be applied to both core and valence states of challenging systems, especially for benchmarking purposes\cite{benzene,SDSRev,QUEST4X}.
Taking iCIPT2 as the solver of CASSCF, we obtain iCISCF(2)\cite{iCISCF} and SOiCISCF(2)\cite{SOiCISCF}, where the symbol `(2)' implies that
the PT2 corrections are only performed within the active space.
iCISCF(2) can handle active spaces as large as (60e,60o), on top of which spin-orbit coupling can further be treated variationally by SOiCISCF(2).

Instead of real applications, a direct comparison of sf-X2C-iCIPT2, SOiCI, and 4C-iCIPT2 (which are based on the sf-X2C, sf-X2C+so-DKH1\cite{X2CSOC1,X2CSOC2}, and
DCB Hamiltonians, respectively) is more relevant in the present context, for they
share the same infrastructure, such that their relative efficiency reflects directly the efficacy of \texttt{MetaWave}.
To this end, the bromine atom is enough for a showcase. The uncontracted cc-pVXZ-DK (X=D, T, Q) basis sets\cite{ccpvxz_br1,ccpvxz_br2}
were used. Keeping the Ar-core frozen, the three basis sets give rise to correlation spaces
(17e, 68o), (17e, 102o), and (17e, 143o), respectively, for both sf-X2C-iCIPT2 and SOiCI, whereas to (17e, 136o), (17e, 204o), and (17e, 286o),
respectively, for 4C-iCIPT2. The orbitals for sf-X2C-iCIPT2 and SOiCI were generated by
a state-averaged CASSCF(5e,3o) calculation, whereas the spinors for 4C-iCIPT2 were generated from the Dirac-Hartree-Fock calculation of  \ce{Br-}
(to avoid symmetry breaking). In the sf-X2C-iCIPT2 calculations, the three components of the ground state $^2P$
are evenly distributed in the $B_{1u}$, $B_{2u}$, and $B_{3u}$ irreps of $D_{2h}$, whereas
in the SOiCI and 4C-iCIPT2 calculations, the use of time reversal symmetry reduces the number of states from 6
($^2P_{j,m_j}$, $j=1/2, 3/2$, $m_j\in[-j, j]$) to 3 ($^2P_{j,|m_j|}$) in the $E_{1/2}$ irrep of $D_{2h}^\ast$.
The so-calculated spin-orbit splittings (SOS) of the $^2P$ state of Br are presented
%in Table \ref{ene_without_3d} when only valence electrons are correlated and
in Table \ref{ene_with_3d}, to be compared with the experimental value\cite{NIST1} of 3685 cm$^{-1}$.
%when the semicore ($3d^{10}$) electrons are further correlated.
%It can be seen that the semicore electrons have a larger influence on the SOS by SOiCI (60$\sim$70 cm$^{-1}$) than that by 4C-iCIPT2 (20$\sim$30 cm$^{-1}$).
%The situation is even more pronounced for heavier elements\cite{SOiCI,4C-iCIPT2}.

In principle, the costs of the ranking, pruning, and PT2 steps of the three methods should be compared separately.
However, the costs for the ranking and pruning steps exhibit considerable variations due to several factors:
(1) the number of iterations, including the macro-cycles of ranking and pruning as well as the micro-cycles of diagonalization, may differ significantly;
(2) the ranking step is governed predominantly by non-computational operations such as memory allocation and sorting;
(3) even the computational operations are memory-bound (meaning that their performance depends on data retrieval from cache/memory)
and cannot fully explore the data locality in practice.
Consequently, the running environment of the computer system
may impact significantly the performance of these operations. As a result, it is not meaningful to compare
the costs for the ranking and pruning steps of different methods.
In contrast, the PT2 step is much less affected by environmental factors. In particular,
it dominates the overall computational cost as the size of the variational space grows.
Therefore, it is more appropriate to compare the PT2 wall times ($T_2$) in the sf-X2C-iCIPT, SOiCI, and 4C-iCIPT2 calculations.
The results are documented in Table \ref{CmprNonRelaRela}, along with the sizes $|P|$ of the variational spaces
for the selection thresholds $C_{\text{min}}=\{5.0,3.0,1.5,0.9,0.5\}\times 10^{-5}$. It is first noted that
the PT2 steps of all the methods scale linearly with respect to the sizes of the variational spaces (cf. Table \ref{ExtraNonRelaRela}).
The PT2 step of SOiCI turns out be about $1.5\sim 2$ times that of sf-X2C-iCIPT2.
This can be understood as follows. If the additional selection of singly excited CSFs by the effective one-body spin-orbit coupling
(SOC) operator (so-DKH1\cite{X2CSOC1,X2CSOC2}) is not to be performed after the selection of sf-X2C-iCIPT2, the PT2 step of SOiCI would precisely double that
of sf-X2C-iCIPT2, due to the doubled size of the variational space (i.e., $2S+1=2$ times $|P|$ of sf-X2C-iCIPT2). In the case of Br, only a small number of quartet CSFs
is brought into the variational space of SOiCI due to the additional SOC selection. Therefore, the fact
that the PT2 step of SOiCI can be less than double that of sf-X2C-iCIPT2 must stem from
the better data vectorization in the case of SOiCI, which
treats the three states simultaneously rather than individually as done by sf-X2C-iCIPT2.
In contrast, the PT2 step of 4C-iCIPT2 is typically $10 \sim 15$ times that of SOiCI, due to changes in the natures of MPBFs and Hamiltonian.

% Table generated by Excel2LaTeX from sheet 'large'
\begin{landscape}
	\begin{table}[htbp]
		\centering
		\caption{Energies (in $E_h$) of $^2P$, $^2P_{3/2}$, and $^2P_{1/2}$ of \ce{Br} Calculated by sf-X2C-iCIPT2, SOiCI, and 4C-iCIPT2
with Uncontracted cc-pVXZ-DK (X=D,T,Q) Basis Sets. %The $3d^{10}4s^24p^5$ Electrons are Correlated.
		}
\begin{threeparttable}
		\begin{tabular}{|c|c|c|cc|cc|cr|}
			\toprule
			\multirow{2}[2]{*}{Basis} & \multirow{2}[2]{*}{$C_{\textrm{min}}$} & sf-X2C-iCIPT2 & \multicolumn{2}{c|}{SOiCI} & \multicolumn{2}{c|}{4C-iCIPT2} & \multicolumn{2}{c|}{SOS\tnote{*}} \\\cline{3-9}
			&
			& $^2P$
			& $^2P_{3/2}$
			& $^2P_{1/2}$
			& $^2P_{3/2}$
			& $^2P_{1/2}$
			& SOiCI & \multicolumn{1}{c|}{4C-iCIPT2} \\
			\midrule
			\multirow{5}[2]{*}{DZ}
			& $5.0\times 10^{-5}$ & -2604.714234  & -2604.722759  & -2604.706763  & -2603.932931  & -2603.916402  & 3511  & \multicolumn{1}{c|}{3628 } \\
			& $3.0\times 10^{-5}$ & -2604.714149  & -2604.722688  & -2604.706685  & -2603.932466  & -2603.916013  & 3512  & \multicolumn{1}{c|}{3611 } \\
			& $1.5\times 10^{-5}$ & -2604.713998  & -2604.722550  & -2604.706540  & -2603.931715  & -2603.915262  & 3514  & \multicolumn{1}{c|}{3611 } \\
			& $9.0\times 10^{-6}$ & -2604.713921  & -2604.722478  & -2604.706465  & -2603.931695  & -2603.915244  & 3514  & \multicolumn{1}{c|}{3611 } \\
			& $5.0\times 10^{-6}$ & -2604.713876  & -2604.722436  & -2604.706421  & -2603.931704  & -2603.915245  & 3515  & \multicolumn{1}{c|}{3612 } \\
			\midrule
			\multirow{5}[2]{*}{TZ}
			& $5.0\times 10^{-5}$ & -2604.882296  & -2604.890842  & -2604.875037  & -2604.101958  & -2604.085561  & 3469  & \multicolumn{1}{c|}{3599 } \\
			& $3.0\times 10^{-5}$ & -2604.882276  & -2604.890839  & -2604.875030  & -2604.098312  & -2604.082083  & 3470  & \multicolumn{1}{c|}{3562 } \\
			& $1.5\times 10^{-5}$ & -2604.882200  & -2604.890781  & -2604.874966  & -2604.098467  & -2604.082223  & 3471  & \multicolumn{1}{c|}{3565 } \\
			& $9.0\times 10^{-6}$ & -2604.882145  & -2604.890738  & -2604.874917  & -2604.098405  & -2604.082176  & 3472  & \multicolumn{1}{c|}{3562 } \\
			& $5.0\times 10^{-6}$ & -2604.882107  & -2604.890711  & -2604.874884  & -2604.098630  & -2604.082379  & 3473  & \multicolumn{1}{c|}{3567 } \\
			\midrule
			\multirow{5}[2]{*}{QZ}
			& $5.0\times 10^{-5}$ & -2604.919564  & -2604.928141  & -2604.912282  & -2604.139351  & -2604.122914  & 3481  & \multicolumn{1}{c|}{3608 } \\
			& $3.0\times 10^{-5}$ & -2604.919758  & -2604.928350  & -2604.912491  & -2604.136693  & -2604.120350  & 3481  & \multicolumn{1}{c|}{3587 } \\
			& $1.5\times 10^{-5}$ & -2604.919916  & -2604.928527  & -2604.912666  & -2604.135511  & -2604.119233  & 3481  & \multicolumn{1}{c|}{3573 } \\
			& $9.0\times 10^{-6}$ & -2604.919954  & -2604.928581  & -2604.912716  & -2604.135646  & -2604.119338  & 3482  & \multicolumn{1}{c|}{3579 } \\
			& $5.0\times 10^{-6}$ & -2604.919963  & -2604.928604  & -2604.912734  & -2604.136107  & -2604.119858  & 3483  & \multicolumn{1}{c|}{3566 } \\
			\bottomrule
		\end{tabular}%
	\begin{tablenotes}
			\item[*] Spin-orbit splitting in cm$^{-1}$ (experimental value\cite{NIST1}: 3685 cm$^{-1}$).
		\end{tablenotes}
\end{threeparttable}
		\label{ene_with_3d}
	\end{table}%
\end{landscape}

\begin{table}[htbp]
	\tiny
	\centering
	\caption{Wall Times$^*$ ($T_2$ in seconds) for PT2 Corrections in
		sf-X2C-iCIPT2, SOiCI, and 4C-iCIPT2 calculations of Br ($3d^{10}4s^2 4p^5$).
	}
	\begin{threeparttable}
		\begin{tabular}{c|c|rr|rr|rr|rr}
			\toprule
			\multirow{2}[2]{*}{Basis set}
			& \multirow{2}[2]{*}{$C_{\text{min}}$}
			& \multicolumn{2}{c|}{sf-X2C-iCIPT2}
			& \multicolumn{2}{c|}{SOiCI}
			& \multicolumn{2}{c|}{4C-iCIPT2}
			& \multicolumn{2}{c}{4C-iCIPT2\tnote{a}} \\\cline{3-10}
			&
			& \multicolumn{1}{c}{$|P|$\tnote{b,c}}
			& \multicolumn{1}{c|}{$T_2$}
			& \multicolumn{1}{c}{$|P|$\tnote{b,d}}
			& \multicolumn{1}{c|}{$T_2$\tnote{e}}
			& \multicolumn{1}{c}{$|P|$\tnote{b}}
			& \multicolumn{1}{c|}{$T_2$\tnote{f}}
			& \multicolumn{1}{c}{$|P|$\tnote{b}}
			& \multicolumn{1}{c}{$T_2$} \\
			\midrule
			\multirow{5}[2]{*}{DZ}
			& $5.0\times 10^{-5}$ & 146949   & 48.9    & 297306   & 139.2 [2.85]  & 66809  & 440.6  [3.16]  & 18148  & 12.5 \\
			& $3.0\times 10^{-5}$ & 266972   & 64.2    & 541068   & 167.7 [2.61]  & 115839 & 759.6  [4.53]  & 28080  & 14.0 \\
			& $1.5\times 10^{-5}$ & 582319   & 112.8   & 1187558  & 245.7 [2.18]  & 267706 & 1716.7 [6.99]  & 50239  & 16.8 \\
			& $9.0\times 10^{-6}$ & 1022725  & 167.6   & 2092972  & 319.0 [1.90]  & 513195 & 3314.9 [10.4]  & 75532  & 20.6 \\
			& $5.0\times 10^{-6}$ & 1906432  & 281.6   & 3915296  & 507.4 [1.80]  & 998099 & 6386.7 [12.6]  & 118985 & 25.5 \\
			\midrule
			\multirow{5}[2]{*}{TZ}
			& $5.0\times 10^{-5}$ & 248384  & 141.6    & 500826   & 286.1 [2.02]   & 115189  & 1380.7  [4.82]  & 57106  & 35.1   \\
			& $3.0\times 10^{-5}$ & 550019  & 249.1    & 1108046  & 425.7 [1.71]   & 247380  & 3999.9  [9.40]  & 103112 & 53.4   \\
			& $1.5\times 10^{-5}$ & 1533813 & 597.6    & 3093278  & 958.8 [1.60]   & 590573  & 12678.1 [13.2]  & 217305 & 94.8   \\
			& $9.0\times 10^{-6}$ & 2967699 & 1040.4   & 5998268  & 1737.4 [1.67]  & 1175598 & 22453.7 [12.9]  & 345719 & 145.5  \\
			& $5.0\times 10^{-6}$ & 5930752 & 2047.4   & 12023362 & 3286.5 [1.61]  & 2504958 & 52175.4 [15.9]  & 563638 & 228.0  \\
			\midrule
			\multirow{5}[2]{*}{QZ}
			& $5.0\times 10^{-5}$ & 313293   & 310.3   & 630388   & 484.6   [1.56]  & 155843  & 3724.6  [7.69]  & 89896   & 92.3    \\
			& $3.0\times 10^{-5}$ & 740258   & 688.9   & 1487852  & 1052.9  [1.53]  & 332023  & 10342.9 [9.82]  & 170270  & 164.0   \\
			& $1.5\times 10^{-5}$ & 2222671  & 1889.9  & 4470458  & 2887.8  [1.53]  & 840020  & 32005.5 [11.1]  & 381604  & 363.3   \\
			& $9.0\times 10^{-6}$ & 4667946  & 3897.4  & 9403060  & 6063.9  [1.56]  & 1686352 & 71487.5 [11.8]  & 674228  & 675.5   \\
			& $5.0\times 10^{-6}$ & 10325265 & 8535.1  & 20843428 & 13209.3 [1.55]  & 3866697 &174537.7 [13.2]  & 1276668 & 1199.8  \\
			\bottomrule
		\end{tabular}%
		\begin{tablenotes}
			\item[*] Performed on a single node with two Hygon 7285 CPUs (32 cores at 2.0 GHz) and 512 GB DDR4 memory.
			\item[a] Only $4s^2 4p^5$ electrons are correlated.
			\item[b] Size of variation space.
			\item[c] Summation of the number of CSFs for each irrep of $D_{2h}$.
			\item[d] Summation of the number of CSFs for each irrep of $D_{2h}$ weighted by spin multiplicity.
            \item[e] Relative cost with respect to sf-X2C-iCIPT2.
            \item[f] Relative cost with respect to SOiCI.
		\end{tablenotes}
	\end{threeparttable}
	\label{CmprNonRelaRela}
\end{table}%

\begin{table}[htbp]
	\centering
	\caption{Linear Fits of PT2 Wall Times versus Sizes of Variational Spaces (see Table \ref{CmprNonRelaRela})
for sf-X2C-iCIPT2, SOiCI and 4C-iCIPT2
calculations of Br ($3d^{10}4s^2 4p^5$).
	}
	\begin{threeparttable}
		\begin{tabular}{c|cl|cl|cl|cl}\toprule
			\multicolumn{1}{c}{\multirow{2}[1]{*}{Basis set}}
			& \multicolumn{2}{|c|}{sf-X2C-iCIPT2}
			& \multicolumn{2}{c|}{SOiCI}
			& \multicolumn{2}{c|}{4C-iCIPT2}
			& \multicolumn{2}{c}{4C-iCIPT2\tnote{a}}
			\\\cline{2-9}
			& Slope \tnote{b}
			& $R^2$
			& Slope \tnote{b}
			& $R^2$
			& Slope \tnote{b}
			& $R^2$
			& Slope \tnote{b}
			& $R^2$
			 \\ \toprule
			DZ    & 1.32  & 0.9991  & 1.00  & 0.998   & 63.9   & 0.99998  & 1.30  & 0.997     \\
			TZ    & 3.34  & 0.9997  & 2.61  & 0.99989 & 211    & 0.998    & 3.81  & 0.9989    \\
			QZ    & 8.20  & 0.99991 & 6.29  & 0.9997  & 445    & 0.99990  & 9.38  & 0.998     \\\bottomrule
		\end{tabular}%
		\begin{tablenotes}
            \item[a] Only $4s^2 4p^5$ electrons are correlated.
			\item[b] Multiplied by $10^4$.
		\end{tablenotes}
	\end{threeparttable}
	\label{ExtraNonRelaRela}
\end{table}%

\section{Conclusion and Outlook} \label{Conclusion}
By utilizing advanced C++ features, such as template metaprogramming and polymorphism,
a highly modular, extendable, and efficient platform, dubbed \texttt{MetaWave}, has been introduced for
unified implementation of many-electron wavefunctions. It is composed of three abstract layers, i.e.,  \texttt{infrastructure},  \texttt{Hamiltonian}, and  \texttt{wavefunction}.
The \texttt{infrastructure} layer supplies unified interfaces for basic linear algebra operations and algorithm templates for parallel computing, abstracting scalar type dependencies and parallel strategies. The  \texttt{Hamiltonian} layer addresses
domain-specific components such as Hamiltonians, MPBFs, symmetries, integrals, and matrix elements, employing a type-trait and tagging system.
The  \texttt{wavefunction} layer implements high-level computational algorithms, benefiting from the abstractions provided by the \texttt{infrastructure} and  \texttt{Hamiltonian} layers.
In this way, Hamiltonians, wavefunctions, and parallel strategies are decoupled to the maximal extent, thereby
reducing code repetition and enhancing code extendability for new functionalities.
Among others, some immediate developments include (1) GPU acceleration of both matrix-vector products and perturbation corrections,
(2) support of more and general Hamiltonians for the extension of \texttt{MetaWave} to a wider range of fermionic and/or bosonic quantum systems,
and (3) support of tensor contractions for unified implementation of relativistic and nonrelativistic coupled-cluster methods.

\section*{Notes}
The authors declare no competing financial interest.

\section*{Acknowledgments}
This work was supported by National Natural Science Foundation of China (Grant No. 22373057) and
Mount Tai Scholar Climbing Project of Shandong Province.

\section*{Data availability}
All data are available in this article.

\newpage

\bibliography{iCI}

\newpage

\noindent TOC graphic

\begin{figure}
	\includegraphics[width=0.7\textwidth]{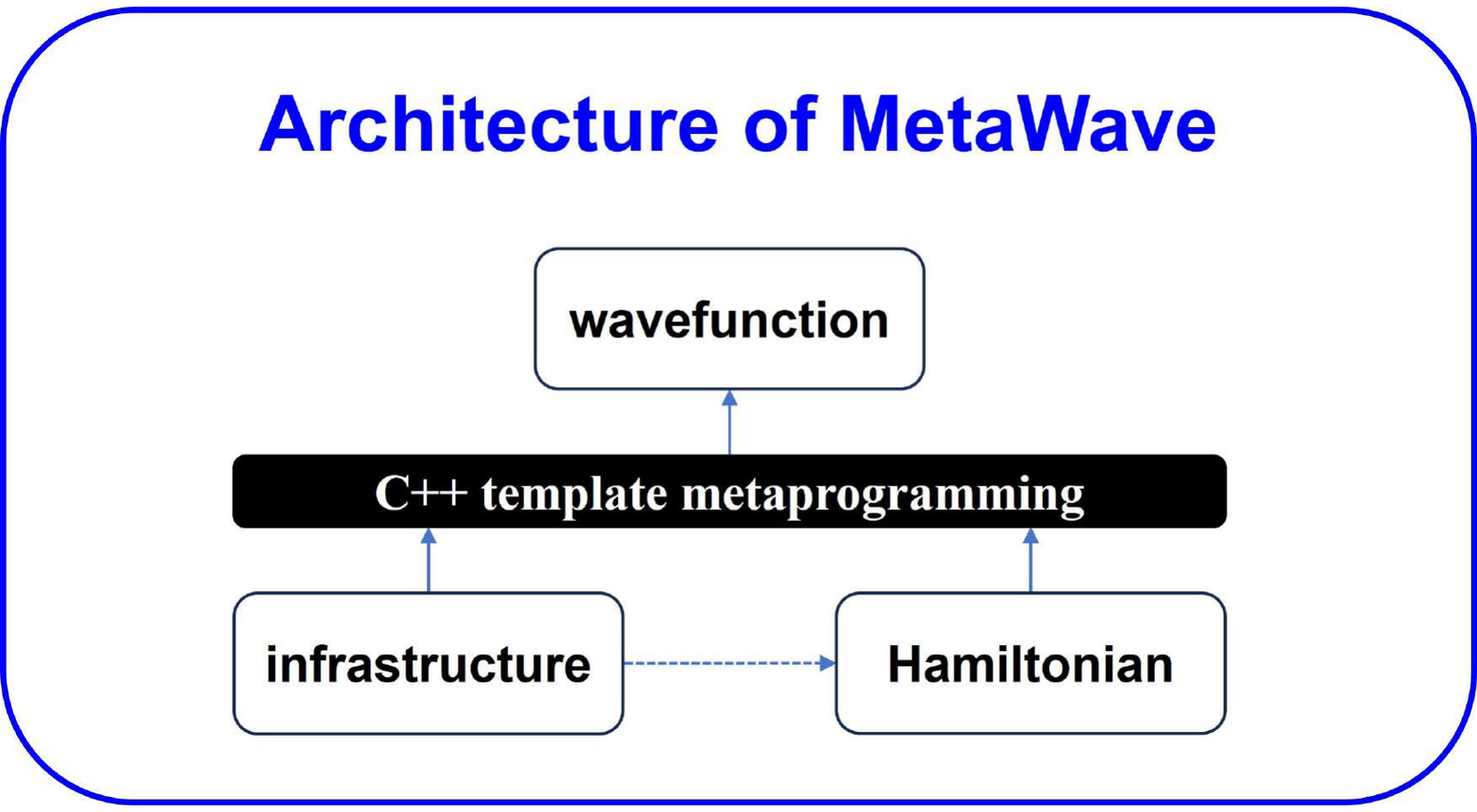}
\end{figure}

\end{document}